\documentclass[aps,prl,floatfix,superscriptaddress,twocolumn,color]{revtex4}
\usepackage{graphicx}
\usepackage{amsmath}
\usepackage{amsfonts}
\usepackage{amssymb}
\usepackage{wasysym}
\usepackage{dsfont}
\usepackage{color}
\definecolor{redmarker}{rgb}{0.9,0.0,0.0}
\definecolor{greenmarker}{rgb}{0.0,0.9,0.0}
\definecolor{bluemarker}{rgb}{0.1,0.1,0.9}
\definecolor{lightblue}{rgb}{0.8,0.8,0.9}

\begin{document}

%%%%%%%%%%%%%%%%%%%%%%%%%%%%%%

%% For titles, only capitalize the first letter
\title{Formation and maintenance of nitrogen fixing cell patterns in filamentous cyanobacteria}

\author{Javier Mu\~noz-Garc\'ia}
\author{Sa\'ul Ares} \email{saul@math.uc3m.es}
\affiliation{Grupo Interdisciplinar de Sistemas Complejos (GISC) and Departamento de Matem\'aticas,
Universidad Carlos III de Madrid, 28911 Leganes, Spain}

%\significancetext{Cyanobacteria produce an important fraction of oxygen on Earth and, together with archaea, fix atmospheric nitrogen used by all other organisms. Some types live in colonies with specialized cells that perform different functions. In particular, the genus Anabaena form filaments in which some cells differentiate forming patterns in order to effectively provide nitrogen for the colony. We present a theory that combines genetic, metabolic and morphological aspects to understand this prokaryotic example of multicellularity. Our results quantitatively reproduce the appearance and dynamics of this pattern and are used to learn how different aspects, like fixed nitrogen diffusion, cell division, or stochasticity, affect it.}

\date{\today}

%%%%%%%%%%%%%%%%%%%%%%%%%%%%%%%%%%%%%%%%%%%%%%%%%%%%%%%%%%%%%%%%computational

\begin{abstract} 
Cyanobacteria forming one-dimensional filaments are paradigmatic model organisms of the transition between unicellular and multicellular living forms.
Under nitrogen limiting conditions, in filaments of the genus {\em Anabaena}, some cells differentiate into heterocysts, which lose the possibility to divide but are able to fix environmental nitrogen for the colony.
These heterocysts form a quasi-regular pattern in the filament, representing a prototype of patterning and morphogenesis in prokaryotes.
Recent years have seen advances in the identification of the molecular mechanism regulating this pattern.
We use this data to build a theory on heterocyst pattern formation, for which
both genetic regulation and the effects of cell division and filament growth are key components.
The theory is based on the interplay of three generic mechanisms:
local autoactivation, early long range inhibition, and late long range inhibition. 
These mechanisms can be identified with the dynamics of {\em hetR}, {\em patS} and {\em hetN} expression.
Our theory reproduces quantitatively the experimental dynamics of pattern formation and maintenance for wild type and mutants. 
We find that {\em hetN} alone is not enough to play the role as the late inhibitory mechanism: a second mechanism, hypothetically the products of nitrogen fixation supplied by heterocysts, must also play a role in late long range inhibition.
%
%We find that {\em hetN} alone is not enough to play the role as the late inhibitory mechanism: the products of nitrogen fixation supplied from heterocysts must also play a role in late long range inhibition.
%
The preponderance of even intervals between heterocysts arises naturally as a result of the interplay between the timescales of  genetic regulation and cell division.
We also find that a purely stochastic initiation of the pattern, without a two-stage process, is enough to reproduce experimental observations.

%the existence of a prepattern derived from a two-stage process is not a necessary requirement to explain any experimental observation.

\end{abstract}

%\pacs{
%05.45.Xt,		%Synchronization; coupled oscillators
%02.30.Ks,			%Delay and functional equations
%87.18.Hf			%Spatiotemporal pattern formation in cellular populations
%87.19.lx			%Development and growth
%}

\maketitle

%% When adding keywords, separate each term with a straight line: |
%\keywords{cyanobacteria| pattern differentiation|prokaryotic development| activator-inhibitor  }
%\keywords{cyanobacteria| pattern formation| activator-inhibitor| heterocyst differentiation| gene regulatory networks  }

%% Optional for entering abbreviations, separate the abbreviation from
%% its definition with a comma, separate each pair with a semicolon:
%% for example:
%% \abbreviations{SAM, self-assembled monolayer; OTS,
%% octadecyltrichlorosilane}

% \abbreviations{}

%% The first letter of the article should be drop cap: \dropcap{}
%\dropcap{I}n this article we study the evolution of ''almost-sharp'' fronts

%% Enter the text of your article beginning here and ending before
%% \begin{acknowledgements}
%% Section head commands for your reference:
%% \section{}
%% \subsection{}
%% \subsubsection{}
\noindent

Cyanobacteria were pioneer organisms to use oxygenic photosynthesis and are currently one of the most successful living groups, occupying a broad range of habitats across all latitudes and producing a large fraction of Earth's photosynthetic activity.
Some types of cyanobacteria form colonies consisting of one-dimensional filaments
%or trichomes
%composed in the presence of fixed nitrogen and the absence of stresses only of vegetative cells carrying photosynthesis.
of vegetative cells carrying photosynthesis.
However, as a response to different environmental stresses, vegetative cells can differentiate into specialized cell types that perform important functions for the survival of the colony. This is a paradigmatic example of a prokaryotic form of life with differentiated cell types.
%In particular vegetative cells may transform into three cell types: heterocysts, akinetes, and hormogonia.
%Here we will focus on the case of heterocyst differentiation. Heterocysts are specialized cells able to fix atmospheric nitrogen into a chemical form usable by vegetative cells.
%

Bacteria and archaea are the only forms of life able to fix atmospheric nitrogen, making them crucial for all living forms on Earth. N$_2$ fixation is catalyzed by nitrogenase, and this enzyme is easily degraded by oxygen.
Some filamentous cyanobacteria have developed a
%an interesting
mechanism to protect nitrogenase from the oxygen produced by vegetative cells.
When external nitrogen sources are scarce, specialized cells called heterocysts appear in a quasi-regular pattern, with intervals of around ten vegetative cells between consecutive heterocysts. %representing a simple example of pattern formation in biological systems.
%Since a continuous outer membrane covers the whole cyanobacterial filament, the fixed nitrogen produced by heterocysts can diffuse inside it and reach vegetative cells.
Since cells can exchange metabolites and small peptides \cite{mariscal07,mullineaux2008,zhang08,flores10,zhang13,omairinasser2014}, the fixed nitrogen produced by heterocysts can reach vegetative cells.
N$_2$ fixation requires high energy consumption. In order to maintain it, nutrients produced by photosynthesis in vegetative cells are shared with heterocysts \cite{flores10,muro-pastor12}.
Upon differentiation, heterocysts lose the possibility to undergo cell division.
%\highlightr{making their fate a sacrifice in favor of the colony}.
However, vegetative cells continue dividing, producing filament growth and increasing the distance between consecutive heterocysts.
As a result new heterocysts differentiate roughly in the middle of the intervals between previously existing heterocysts. This dynamic process of differentiation allows the overall pattern to conserve its properties over time.

The biology of heterocyst formation has been the subject of intensive work \cite{flores10,kumar10}.
%, see Refs. \cite{flores10,kumar10} for recent reviews.
%
Most studies focus on the strain PCC 7120 of the genus {\it Anabaena},
%(also known as Nostoc sp. PCC 7120),
%
%The abbreviation "sp." is used when the actual specific name cannot or need not be specified. The abbreviation "spp." (plural) indicates "several species". These abbreviations are not italicised (or underlined).[44] For example: "Canis sp." means "an unspecified species of the genus Canis", while "Canis spp." means "two or more species of the genus Canis".
%
%
which has become a model organism in the field. Recent quantitative
experimental work has produced a wealth of data on vegetative cell intervals between heterocysts under a number of mutations and experimental conditions. %\cite{haselkorn98,yoon98,yoon01,golden03,khudyakov04,wu04,borthakur05,orozco06,risser07,risser09,higa10,feldmann11}.
Moreover, the one-dimensional nature of this pattern forming
system represents a very appealing system for theoretical and
mathematical modeling
\cite{wilcox73,wolk75,meinhardt08,pinzon06,allard07,gerdtzen09,zhu10,brown2012a,brown2012b,brown2014,torres2015,ishihara2015}.\\

%
%Some efforts have been made recently in this line  to explain aspects of heterocyst differentiation %
%
%For example Meinhardt proposed a qualitatively reaction-diffusion model \cite{meinhardt08}, where the transcription factor HetR activates differentiation, and the diffusible protein PatS inhibits it. Pinzon and Ju \cite{pinzon06} estimated the yield of nitrogen produced under different growth conditions using a method akin to a mean-field without considering the spatial extension. Allard et al. studied the effects of cell division and nitrogen distribution in the filament, without incorporating details of the genetic regulation \cite{allard07}. Gerdtzen et al. proposed a bioinformatics model for the formation of heterocysts patterns, based on the iteration of matrix operators that determine the evolution of the state of a cell \cite{gerdtzen09}, and Zhu et al. proposed a one-dimensional continuum model based on partial differential equations with moving boundary conditions to model the pattern appearance \cite{zhu10}. 

%
%\begin{tcolorbox}[width=\linewidth,colback={lightblue},title={Significance},colbacktitle=lightblue,coltitle=black]    
\noindent\fbox{%
    \parbox{0.5\textwidth}{%
{\bf Significance}\\ Cyanobacteria produce an important fraction of oxygen on Earth and, together with archaea, fix atmospheric nitrogen used by all other organisms. Some types live in colonies with specialized cells that perform different functions. In particular, the genus Anabaena forms filaments in which some cells differentiate, forming patterns in order to effectively provide nitrogen for the colony. We present a theory that combines genetic, metabolic, and morphological aspects to understand this prokaryotic example of multicellularity. Our results quantitatively reproduce the appearance and dynamics of this pattern and are used to learn how different aspects, like fixed nitrogen diffusion, cell division, or stochasticity, affect it.
   }%
}
%\end{tcolorbox} 
%
\newpage

Despite these efforts, many processes and genetic mechanisms involved in the regulation of heterocyst differentiation, pattern formation, and maintenance remain poorly understood.
%
%unknown or its function is not clear, and many open question related to the pattern appearance and its maintenance.
For example, it has not been clarified which particular vegetative cells differentiate into heterocysts and if this is related to some inherited predisposition \cite{asai09,toyoshima10}. Other open questions are whether the differentiating cells are selected stochastically,  how  the typical spacing between heterocysts is determined for %\textit{de novo}
cells differentiating at early and late times, or which mechanisms induce the appearance of multiple contiguous heterocysts, the so called Mch phenotype \cite{yoon98}, in some mutants. 

%%
%In this work we formulate a basic theoretical description of heterocyst pattern formation that includes the main genetic regulations identified in the process.
%%
%In particular we focus on the interactions between the {\it hetR}, {\it patS} and {\it hetN} genes, since they have been shown to be essential to control the formation of the initial pattern and its maintenance.
%%
%We also explicitly consider the products of nitrogen fixation supplied by heterocysts.
%%
%The effects of other genes and metabolic processes are taken into account in an effective way.
%%
%%
%We make use of recent biochemical data to propose a reaction model which is set in the framework of a growing filament, taking into account the stochastic nature of cell division times.
%%
%This model quantitatively reproduces the main features of the differentiation process for wild type and mutants, including the appearance and maintenance of a quasi-regular pattern of heterocysts.
%

\subsection{Basic genetics of heterocyst differentiation} \label{genetics}
%\highlightr{Move to intro?}
%\highlightr{HetR binds DNA as a dimer: \cite{kim2011,kim2013} and can even form tetramers \cite{kim2013}.
%
%Genes involved in early heterocyst differentiation, like {\em hetZ}, have transcription from their promoter correlated to the expression level of {\em hetR} and inhibition by RGSGR \cite{du2012}.
%}
There is a large number of processes involved in the regulation of heterocyst pattern formation. In addition to nitrogen levels and other environmental aspects, many genes and factors play a role \cite{ehira2013}.
When nitrogen stress is perceived, the transcriptional regulator \textit{ntcA} is important to trigger heterocyst differentiation \cite{herrero04,zhang06}, directly or indirectly controlling the expression of several genes \cite{valladares08,gonzalez2013} including \textit{hetR}.
The gene \textit{hetR} is central to heterocyst differentiation. Its expression is the main positive regulatory factor in heterocyst development \cite{buikema91,black93,haselkorn98,risser07}. The expressions of \textit{ntcA} and \textit{hetR} are mutually dependent,
%but in contrast to the former,
and the latter seems to be necessary and sufficient for heterocyst development, even under conditions of excess of external nitrogen \cite{buikema01}.
%
%an autoregulatory gene central to heterocyst differentiation since its expression seems to be the main positive regulatory factor in heterocyst development \cite{buikema91,black93}. \textit{ntcA} and \textit{hetR} expressions are mutually dependent, but in contrast to the former, the later seems to be exclusively required and sufficient for heterocyst development, even under conditions of excess of external nitrogen as shown in Ref. \cite{buikema01} where HetR is artificially drive by the copper-inducible \textit{pet}E promoter.
%
Thus, positive autoregulation of \textit{hetR} is required for differentiation and is particularly significant in developing heterocysts \cite{black93,rajagopalan10}.
In addition to \textit{ntcA}, other genes such as \textit{patA}, \textit{hetF} and \textit{hetP} also regulate heterocyst differentiation \cite{liang92,wong01,risser08,higa10}.
%
%however in order to elucidate the main mechanisms controlling heterocyst pattern establishment we will focus on other gene products that have been proved to be essential for controlling the pattern formation. 

The gene \textit{patS} is a negative regulator of \textit{hetR} that suppresses differentiation when overexpressed and induces a Mch phenotype when deleted \cite{yoon98,yoon01,golden03}.
PatS is a short peptide, predicted to be 13 or 17 amino acids, containing a carboxy-terminal pentapeptide RGSGR
%known as PatS-5
that prevents DNA binding activity of HetR \cite{feldmann11,huang04} and inhibits differentiation when added to culture medium \cite{yoon98}.
The expression of \textit{patS} in small groups of vegetative cells was shown to diminish the levels of HetR in adjacent cells \cite{risser09} suggesting that a PatS-dependent signal can diffuse along the filament \cite{corrales-guerrero2013}.
%since it is believed that it needs to be processed when transported from cell to cell to function properly \cite{wu04}.
%
It has been observed that \textit{patS} is strongly expressed in developing heterocysts \cite{yoon98,yoon01} coming back to low levels after 24 hours of nitrogen deprivation. 
%
%Overexpression of this gene results in a complete suppression of heterocyst development and its 
%
Although lack of \textit{patS} expression initially produces a pattern with frequent contiguous heterocysts and short intervals between separate ones, later this pattern
tends to become more similar to a wild-type pattern\cite{yoon01}, suggesting the presence of other patterning signals.

Thus, \textit{hetR} and \textit{patS} could be enough to obtain a minimalistic understanding of heterocyst pattern formation at early stages. However, other players have to be taken into account to explain pattern maintenance after an initial pattern of differentiated heterocysts appear.
The key player for this role is the inhibitory factor \textit{hetN}. Akin to PatS, HetN also contains a RGSGR motif, raising the possibility that a RGSGR-containing peptide derived from the full protein diffuses from cell to cell \cite{higa12}.
In contrast to \textit{patS} mutants, \textit{hetN} mutants have a differentiation pattern similar to the wild type at the initial stages of nitrogen depletion and a Mch phenotype after 48 hours \cite{callahan01}, suggesting that \textit{hetN} expression is activated later than that of \textit{patS}. When both genes are suppressed, almost all cells along the filament eventually differentiate causing lethal levels of heterocysts \cite{borthakur05}.

Metabolites have also been suggested to play a role in filament patterning, in particular the fixed nitrogen products produced by heterocysts. It has been suggested that they inhibit heterocyst differentiation \cite{yoon01}, although experiments with {\em Anabaena variabilis} \cite{thiel01,thiel04} did not find an observable effect.

%The question remains open because filament growth is impaired in mutants in which heterocysts cannot supply fixed nitrogen \cite{kumar10}, making experiments infeasible. Thus, a theoretical approach is an invaluable tool to examine this problem.

%\highlightr{Dar m\'as detalles de los mecanismos y la interacci\'on entre genes (decir que HetR es un dimero <- Hecho m‡s adelante, al describir la Fig. 1).\\
%
%HetR binds DNA as a dimer: \cite{kim2011,kim2013} and can even form tetramers \cite{kim2013}.
%
%Genes involved in early heterocyst differentiation, like {\em hetZ}, have transcription from their promoter correlated to the expression level of {\em hetR} and inhibition by %RGSGR \cite{du2012}.
%}

%%%%%%%%%%%%% Figure NETWORK 1 %%%%%%%%%%%%%%%%%%%%%%%
\begin{figure}[t]
\begin{center}
\includegraphics[width=\linewidth]{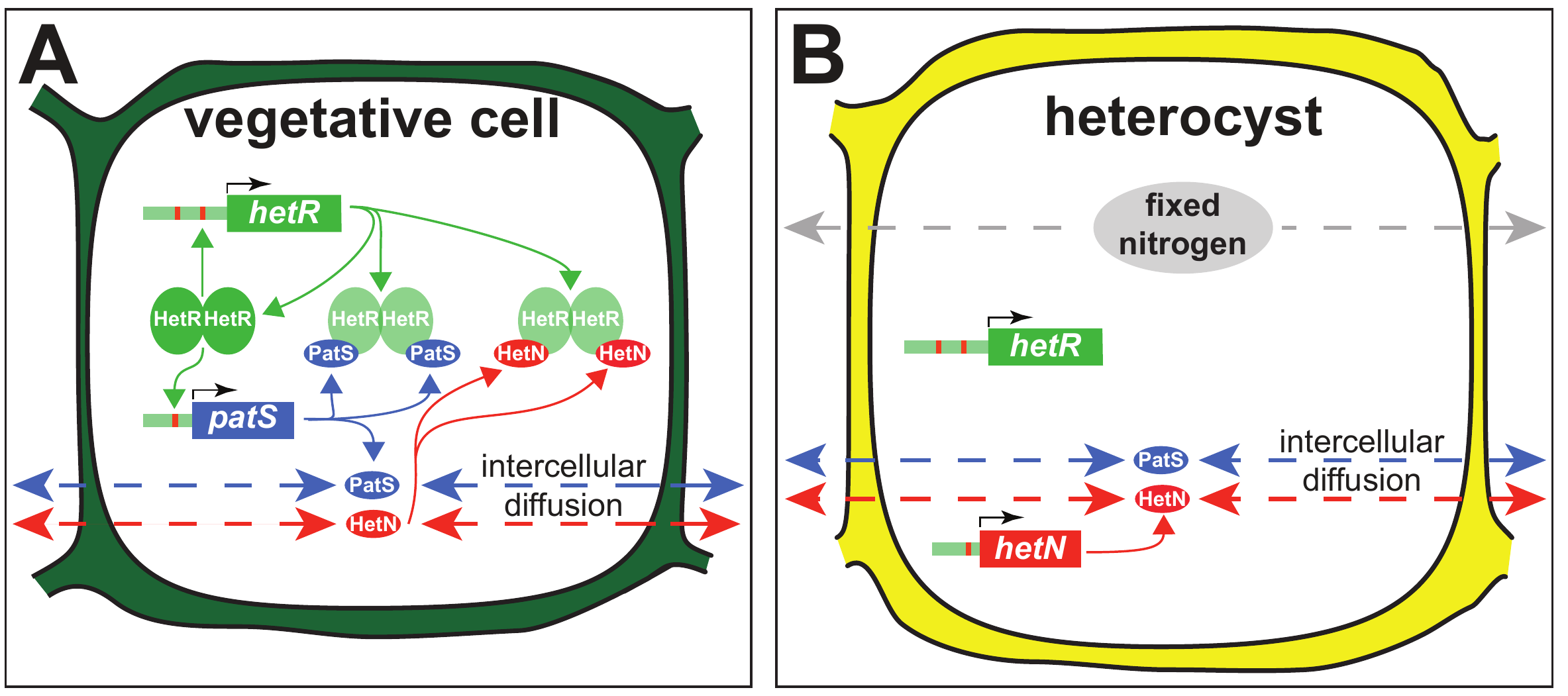}
\end{center}
\caption{Minimal model of the genetic network. ($A$) In vegetative cells HetR dimers can activate the expression of \textit{hetR} and \textit{patS}. ($B$) In heterocysts \textit{hetN} is expressed constitutively and fixed nitrogen products are produced. Active products of PatS and HetN, possibly the RGSGR pentapeptide, can diffuse between cells of any kind in the filament and bind HetR, preventing it from binding DNA. Fixed nitrogen products also diffuse to other cells and contribute weakly to inhibit differentiation.
\label{Fig.model}}  
\end{figure}
%%%%%%%%%%%%%%%%%%%%%%%%%%%%%%%%%%%%%%%%%%%%

\section{Results} \label{results}

\subsection{Minimal model for the heterocyst differentiation gene regulatory network} \label{model}

Although the chemical and genetic regulation of heterocyst differentiation involves a great number of factors \cite{flores10,kumar10,ehira2013}, we will focus on the core 
mechanisms involved: local autoactivation, early long range inhibition, and late long range inhibition. For concreteness, we will identify these generic mechanisms with the dynamics of three genes: \textit{hetR}, \textit{patS} and \textit{hetN}, and use what is known of their biochemistry to build a theoretical model.
%
%The effects of the levels of N$_2$ or other genes like \textit{ntcA} or \textit{patA} can be  understood to first approximation to be included in a phenomenological way as %factors affecting the parameters of our model.
To first approximation, the effects of other genes like \textit{ntcA} or \textit{patA} could be included in a phenomenological way, as factors affecting the  parameters.
This approach will not capture all the subtleties that the explicit inclusion of all possible effects would produce, but has the advantage of being generic, clear and simple, allowing a more systematic analysis of the model.
%offering more insightful conclusions.
%
%For instance, we have ignored the effect on cell growth of initial nitrogen deprivation
%
%Three main mechanisms: local autoactivation (\textit{hetR} ), early long range inhibition (\textit{patS}), and late long range inhibition (\textit{hetN}).  
%
%
%Moreover, the dynamics of the interactions between factors left out of our model are still poorly understood, which would involve heavy speculation to include them.

Fig.~\ref{Fig.model} shows a diagram of the minimal genetic network considered in our theoretical description. Upon nitrogen deprivation, \textit{hetR} is expressed, partly in a constitutive way \cite{rajagopalan10}. Its protein acts as a dimer \cite{huang04,kim2011,kim2013}, binding the \textit{hetR} and \textit{patS} promoter regions and activating expression.
HetR has recently been observed also as a tetramer \cite{valladares15}; including that information in our model would only change parameter values but not the dynamics of differentiation.
A PatS product, the RGSGR pentapeptide, can bind HetR with a 1:1 stoichiometry, and RGSGR-tagged HetR molecules can no longer bind DNA \cite{feldmann11,hu2015}. 
%
%Hence, inhibition of HetR activity by PatS would be due to sequestration or titration of HetR dimers 
%
%
Since HetN also contains the RGSGR motif, we assume that it binds and affects HetR analogously to PatS. PatS is only produced in vegetative cells, while HetN is exclusively produced in heterocysts. Since expression of HetR remains high in heterocysts \cite{zhou98}, we model it through constitutive expression.
We will show later that a second late inhibitor with fast diffusion and a weaker inhibition power than HetN is required to reproduce some experimental findings. A likely candidate for this inhibitor could be the effect of fixed nitrogen products.
Finally, we include intercellular diffusion of PatS, HetN, and fixed nitrogen products.
%Although for simplicity we refer to the species in our model as PatS and HetN, more rigorously they can understood as their active products, with the RGSGR pentapeptide a prominent candidate.
%
For simplicity we do not represent RGSGR explicitly in our model, making instead its effect proportional to the concentration of PatS or HetN.
Vegetative cells that maintain a threshold level of HetR during a predefined period of time are switched to heterocyst status.
%In a generic way, the model represents the interplay of three mechanisms:
%local autoactivation (\textit{hetR}), early long range inhibition (\textit{patS}), and late long range inhibition (\textit{hetN}). 

In contrast to static patterns that are permanently defined once formed, the heterocyst pattern is a dynamic one: vegetative cells keep on growing and dividing, causing the intervals between contiguous heterocysts to continuously increase in size, until eventually a cell of the interval differentiates into a new heterocyst. Thus, we have included cell growth and division as a key ingredient of this model. Moreover, the stochastic nature of gene expression cannot be ignored, and especially the noise in the duration of the cell cycle can play a relevant role in defining the statistics of vegetative intervals between contiguous heterocysts. For this reason noise should be included in a theoretical description, both in the genetic network and in cell division.
The details of the mathematical formulation, analysis and computer simulation of our model are described in the {\em Supporting Information}.

From a physical perspective, the local (non-diffusive) character of the activator, the distinction between inhibitors acting at different times, and the dynamic character of the growing filament, are the features that set this problem apart from other reaction-diffusion  pattern forming systems.

%%%%%%%%%%%%% Figure WT filament %%%%%%%%%%%%%%%%%%%%%%%
\begin{figure}[t]
\begin{center}
\includegraphics[width=\linewidth]{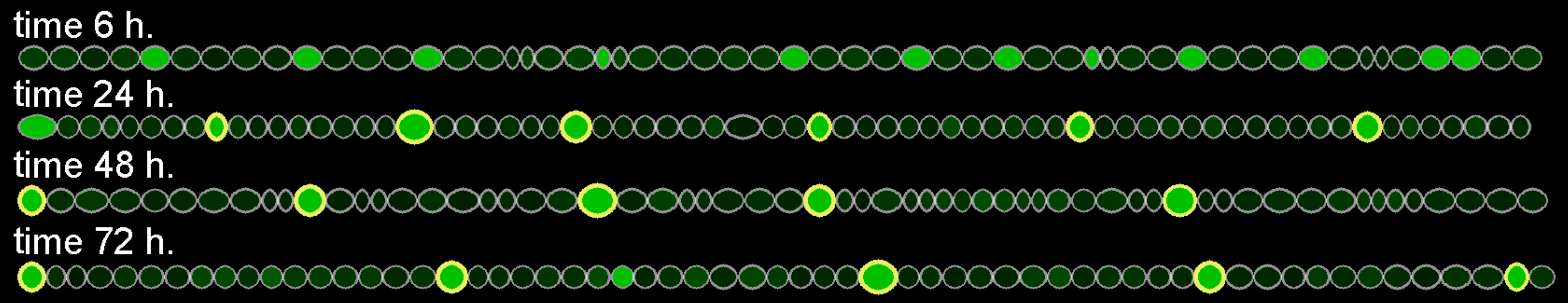}
\end{center}
\caption{Temporal evolution of a filament for the wild type. Heterocyst cells have a yellow membrane. The intensity of the green color shows the level of HetR concentration.
%
%Cells at the left border of the figure are neighbors if the cell at the right border of the row immediately below.
%
Length of the cells represents its value in the simulation.
See also Fig.~S1 and Movie~S1.
\label{Fig.WT}}
\end{figure}
%%%%%%%%%%%%%%%%%%%%%%%%%%%%%%%%%%%%%%%%%%%%%%%%%

\subsection{The theoretical model quantitatively reproduces wild-type patterns}

A first feel of the properties of our model can be obtained by visual inspection of the patterns it produces (Fig.~\ref{Fig.WT}, Fig.~S1 and Movie~S1). Cells with high levels of HetR appear early \cite{jang2009}, frequently in clusters. Eventually only some isolated cells differentiate into heterocysts, while the others revert to a low HetR state.
%
%Frequently cells with high HetR levels are clustered together, but only one of them eventually differentiates into a heterocyst. The HetR level of its neighbors decreases some minutes before or after this differentiation. The cause is that vegetative cells with large levels of HetR are the main synthesizers of the diffusible inhibitor PatS, sequestering HetR molecules and disrupting the positive feedback loop of HetR production.
%%preventing neighbor cells from differentiating.
%
Once a heterocyst pattern is established, cell division increases the vegetative interval length between heterocysts. At some point, one or a few cells roughly in the middle of the interval increase their HetR levels and, finally, just one vegetative cell differentiates into a new heterocyst, thus maintaining the quasi-regularity of the pattern.

\begin{figure}[t]
\begin{center}
\includegraphics[width=\linewidth]{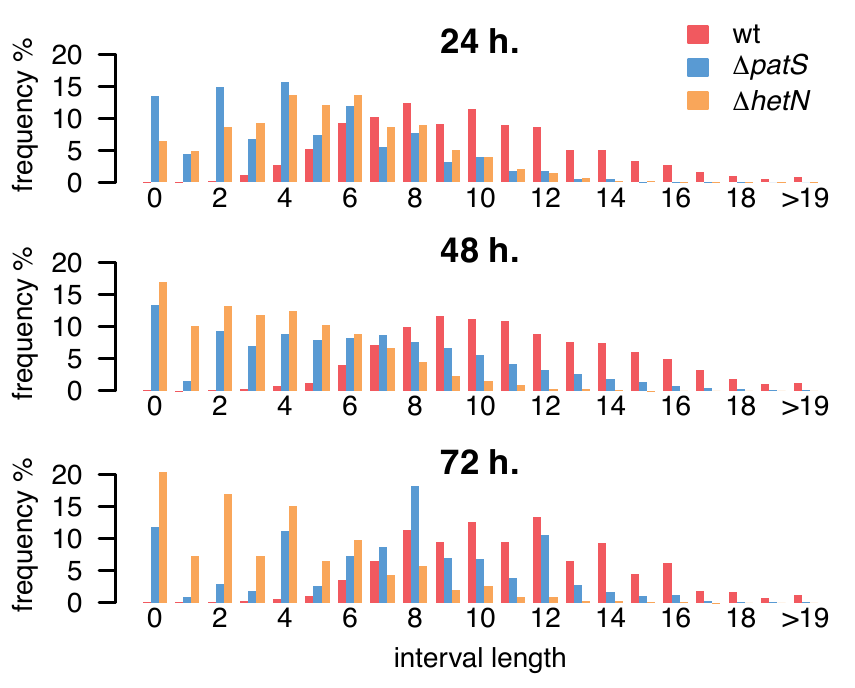}
\end{center}
\caption{Histograms of the number of vegetative cells between heterocysts. Conditions and times after nitrogen deprivation as indicated.
%Histograms determined averaging over 192 different filaments of 150 initial vegetative cells.
%\highlightr{Axis label: Number of contiguous vegetative cells. COMBINE ALL PHENOTYPES IN A SINGLE FIGURE USING R.}
%Interval lengths of greater than or equal to 23 are shown as 23. 
\label{Fig.Histogram}}
\end{figure}

The properties of heterocyst patterns can be quantified making a histogram showing the relative frequency of vegetative intervals of each length. %\cite{yoon98,yoon01,khudyakov04,wu04,borthakur05,risser09,corrales-guerrero2013,meeks02,young10,risser2012}.
%
%Also some interesting data in \cite{asai09}. Pretty Fig.\ 3 in \cite{rajagopalan10}, allowing to estimate some timescales.
%
%No histogram but average interval \cite{orozco06}.
%
Changes in the shape of this histogram allow to characterize differences between different phenotypes, and also between different stages of filament growth after nitrogen deprivation.
We have obtained these histograms and other statistical data averaging over 192 different filaments of 150 initial vegetative cells.
The histograms in Fig.~\ref{Fig.Histogram} are remarkably similar to experimental ones \cite{yoon98,yoon01,khudyakov04,wu04,borthakur05,risser09,corrales-guerrero2013,meeks02,young10,risser2012}, reproducing details such as the increase of the typical length of vegetative intervals with time, the gradual broadening of the shortest intervals observed on the filament, or the higher frequency of even intervals.
The analysis can be pushed further from  qualitative observation to quantitative characterization.
This can be done using successive moments of the distributions given by the histograms: mean, variance and skewness. The mean gives an estimation of the average interval length, and the variance, of the width of the intervals distribution. The skewness is a measure of the asymmetry of the distribution. Positive skewness indicates that the right tail of the distribution is heavier than the left tail, which is the situation generally found in experiments.
%As an alternative to the mean interval length, some publications show the average percentage of heterocysts in the filament.
%
We have compared the temporal evolution of these magnitudes in our simulations together with equivalent statistics extracted from experimental histograms in the literature (Fig.~\ref{Fig.Results}).
%for wild type and \textit{patS} mutant.
%
A first observation for the wild type is that the mean interval length
slightly increases from 10 cells at 24 hours up to 12 cells after a few days.% Fig.~\ref{Fig.Results}A. 

%%%%%%%%%%%%% Figure histograms %%%%%%%%%%%%%%%%%%%%%%%

%%%%%%%%%%%%%%%%%%%%%%%%%%%%%%%%%%%%%%%%%%%%%%%%%

In the following we proceed to discuss other conditions.
To start with,
in experiments all cells of \textit{hetR} mutants remain in the vegetative state \cite{orozco06}. In the same way, since HetR is necessary for differentiation in our model, no heterocysts form in that condition.% in our model.
The same is true for any double knockdown involving \textit{hetR}.
Overexpression of \textit{hetR} induces differentiation of all the cells in the filament.
 %the cells in the filament grow and divide remaining vegetative (not shown). Note that in a more realistic description, cells would have their growth impaired under nitrogen deprivation, and growth would only resume in wild type filaments upon nitrogen fixation by newly formed heterocysts \cite{brown2014}. These processes are out of the scope of our model.

\subsection{PatS promotes early pattern formation}

\begin{figure}[b]
\begin{center}
\includegraphics[width=\linewidth]{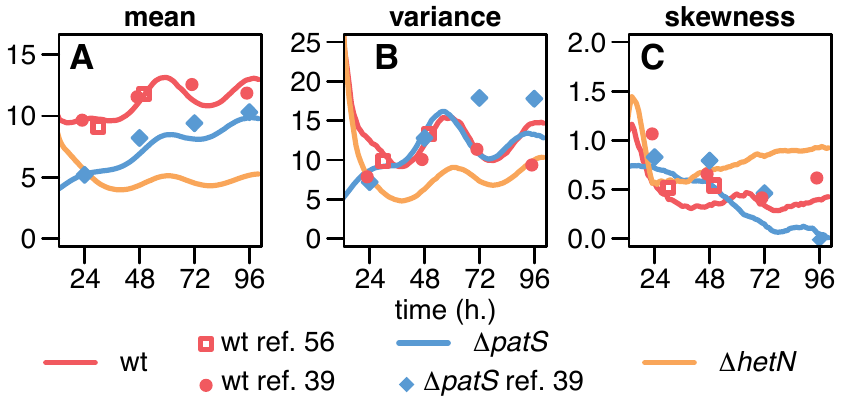}
\end{center}
%\caption{Quantitative dynamics of heterocyst patterning.
\caption{Dynamics of heterocyst patterning.
%for wild type, \textit{patS} and \textit{hetN} mutants in our model compared to experimental results. 
Time evolution of the ($A$) mean, ($B$) variance, and ($C$) skewness of the interval length distribution for our model (lines) and experimental results (symbols), as indicated.  
%$\Delta${\em hetN:}$\Delta$N$_2$ denotes a condition with no late long range inhibition of any kind: no inhibition by HetN or fixed nitrogen products.
See Fig.~S4A for a comparison of theory and experimental data on {\em hetN} loss of function. %\highlightr{[Update ref. [6] to \cite{yoon01} and ref. [8] to \cite{khudyakov04}.]}
\label{Fig.Results}}
\end{figure}

\begin{figure}[t]
\begin{center}
\includegraphics[width=\linewidth]{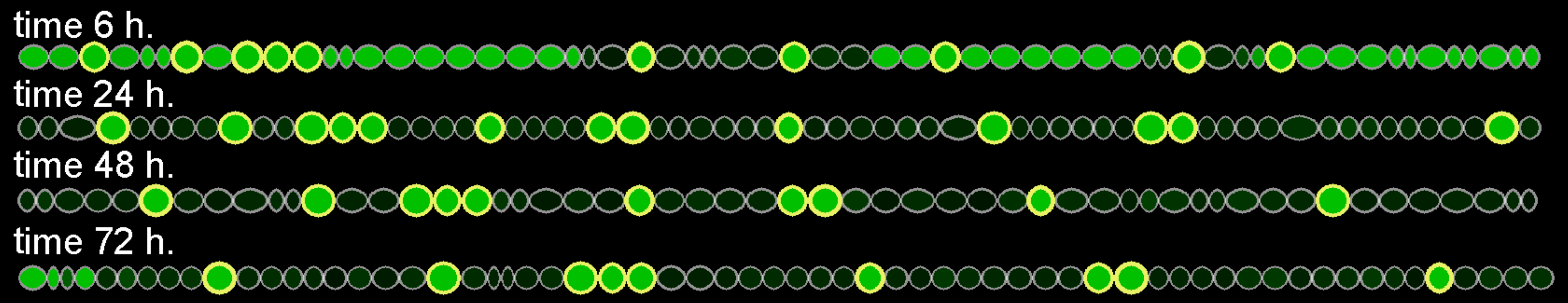}
\end{center}
\caption{Temporal evolution of a filament for the \textit{patS} mutant. Details as in Fig.~\ref{Fig.WT}.
See also Fig.~S2 and Movie~S2.
\label{Fig.APatS}}
\end{figure}

The early behavior of the \textit{patS} mutant is quite different from the wild type (Fig.~\ref{Fig.APatS}, Fig.~S2, and Movie~S2). Due to the lack of PatS, prior to differentiation there is no sequestration of HetR that can weaken its autoinducing feedback loop, so all cells build up high levels of HetR at short times. The first cells to differentiate start producing HetN that diffuses to their neighbors and prevents others from differentiating; however, an abnormally high number of cells, some of them contiguous, have already become heterocysts. This is the Mch phenotype observed in experiments \cite{yoon98, yoon01, borthakur05, orozco06, kumar10}. This high number of heterocysts inhibit differentiation until the vegetative interval between them becomes large enough. For this reason, at long times the Mch phenotype is attenuated and the pattern becomes more similar to wild type. Since new cells cannot appear between contiguous heterocysts, this phenotype persists as a hallmark of the irregular initiation of pattern formation in \textit{patS} mutants.
The histograms for the \textit{patS} mutant (Fig.~\ref{Fig.Histogram})
show the predominance of contiguous heterocysts and short intervals at early times. 
%As time progresses, the histograms become more similar to those for the wild type, except for the persistence of the contiguous heterocysts.
%
%shows that
Starting from mean intervals of 5 cells at 24 h., the mean converges to values similar to wild type after 4 d, as in experiments (Fig.~\ref{Fig.Results}A).

These results confirm that PatS is essential during the first stages of pattern induction to produce a regular pattern. At late stages a regular pattern can be maintained even in the absence of PatS, pointing to the need of a subsequent inhibitor key for pattern maintenance. Continuing the study of the effect of PatS on filament development, we have also simulated PatS overexpression. As observed in experiments \cite{yoon98}, we found a complete repression of heterocyst differentiation in the filament.

%%%%%%%%%%%%% Figure mean, variance, skewness %%%%%%%%%%%%%%%%%%%

%%%%%%%%%%%%%%%%%%%%%%%%%%%%%%%%%%%%%%%%%%%%%%%%%

\subsection{HetN is necessary, but not sufficient, for pattern maintenance: a second inhibitor, probably fixed nitrogen products, also plays a role}

\begin{figure}[b]
\begin{center}
\includegraphics[width=\linewidth]{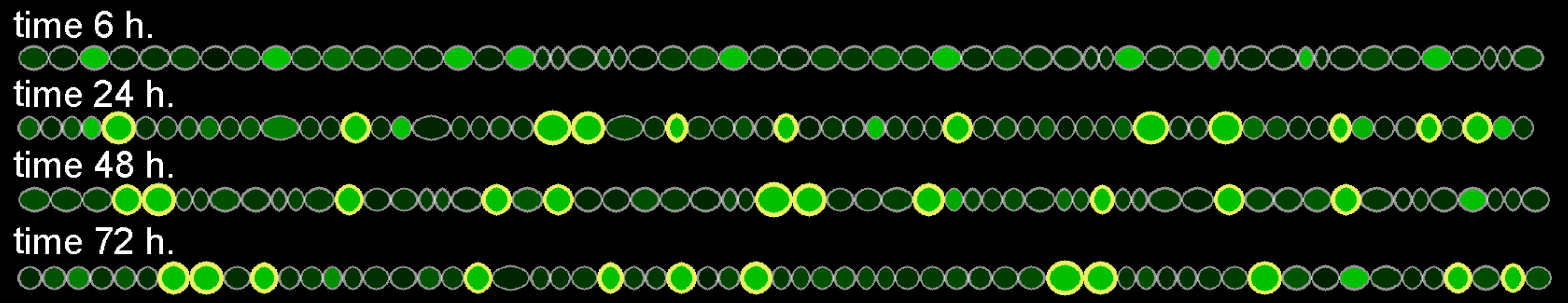}
\end{center}
\caption{Temporal evolution of a filament for the \textit{hetN} mutant. Details as in Fig.~\ref{Fig.WT}.
See also Fig.~S3 and Movie~S3.
\label{Fig.AHetN}}
\end{figure}

In the case of \textit{hetN} loss of function, the presence of PatS assures a correct pattern formation until a first generation of heterocysts differentiates (Fig.~\ref{Fig.AHetN}, Fig.~S3, and Movie~S3). PatS produced in protoheterocysts inhibits the \textit{hetR} feedback loop, but once those protoheterocysts differentiate into heterocysts, they cannot inhibit differentiation in their neighbors since they do not produce HetN.
If the only mechanism of late long range inhibition in the model is the \textit{hetN} gene, in its absence, once a first pattern is established more and more cells continue to differentiate, until eventually the filament is almost entirely composed of heterocysts (Fig.~S4A and Movie~S4).
%Fig.~\ref{Fig.Results}A.
This is far from experimental observations, where the loss of \textit{hetN} condition does induce an increase in the number of heterocysts and a Mch phenotype that eventually stabilizes with time, forming a pattern with roughly 20\% of heterocysts \cite{callahan01,borthakur05,orozco06,corrales-guerrero2014,corrales-guerrero2014plos}.
This hints that a second late long range inhibition mechanism must be at play to avoid the fatal differentiation of the whole filament. A natural candidate is the effect of fixed nitrogen products produced by heterocysts \cite{yoon01,kumar10}, although in principle it could be some other genetic or metabolic species. We have included it in our model explicitly, using a mechanism similar to that of HetN but with weaker inhibition power and larger diffusivity, which accounts for the smaller molecules involved (a detailed theoretical description of the effect of fixed nitrogen can be found in \cite{torres2015}). When the inhibition by fixed nitrogen products is taken into account, the model reproduces quantitatively the experimental data from \textit{hetN} loss of function (Fig.~S4A).
%Fig.~\ref{Fig.Results}A.
This suggests that in a {\em hetN} loss of function background, fixed nitrogen plays a non-negligible role in the maintenance of the pattern, a question that has been difficult to assess experimentally \cite{thiel04,kumar10}. 
HetN and fixed nitrogen products are not required for the appearance of the pattern, but are essential for its maintenance at late stages.
Since {\em hetN} is a much stronger inhibitor in our model, filaments in which fixed nitrogen products have no function are identical to wild type (Fig.~S5 and Movie~S5), in agreement with experiments in {\em Anabaena variabilis} \cite{thiel01,thiel04}. 
%It would be interesting if these experiments could be repeated in a {\em hetN} background. Massive differentiation would indicate that fixed nitrogen products are indeed the second late inhibitor predicted by our theory.
%YO QUITARÍA LA FRASE ANTERIOR. DEBERIA IR ANTES DONDE SE HABLA DEL DOBLE MUTANTE, PERO YA SE DICE EL RESULTADO QUE OBTENEMOS. EL CLARO QUE SERIA INTERESANTE COMPROBARLO EXPERIMENTALMENTE 

As in experiments \cite{callahan01}, overexpression of HetN in our simulations prevents heterocyst differentiation. 
%The result is that at late times the heterocyst pattern is disrupted, with increasing over-differentiation and appearance of a severe Mch phenotype \cite{callahan01, borthakur05, orozco06}. Histograms clearly show how the early phenotype relatively similar to wild type becomes increasingly disrupted over time, Fig.~\ref{Fig.Histogram}.
%
%\highlightr{[SUPPLEMENTARY??? ORDER OF FIGURES!]}.

We have also simulated the double \textit{patS}:\textit{hetN} knockdown condition. 
Since in our model only fixed nitrogen products are working as a weak inhibitor, most of the cells in the filament quickly differentiate into heterocysts, as experimentally observed \cite{borthakur05}, although as late as after 100 hours there is still a small fraction of vegetative cells. Under PatS or HetN overexpression on a \textit{hetN} or \textit{patS} mutant background, respectively, no heterocysts form, as observed \cite{borthakur05}.

Variance and skewness from our simulations also quantitatively compare well with experimental data (Fig.~\ref{Fig.Results}B and C).
%
%\highlightr{
%In the \textit{hetN} mutant the variance is smaller because the increasing Mch phenotype causes small intervals to dominate (Fig.~\ref{Fig.AHetN}), in agreement with reported low mean standard errors \cite{corrales-guerrero2014,corrales-guerrero2014plos}. This does not mean that long intervals do not appear: we have observed intervals as long as 20 cells.
%
%The skewness, however, increases over time, since the peak of the distribution at long times is at zero. Note that these behaviors are even stronger if the effect of fixed nitrogen products is not taken into account.}
%
%CREO QUE QUITARIA EL PÁRRAFO ANTERIOR. QUITAMOS LOS CONTIGUOS ASÍ QUE NO ERA CIERTO LO QUE DECIAMOS. LO DEJARIA SIMPLEMENTE ASI. NO CREO QUE HAGA FALTA JUSTIFICARLO, PERO PODRÍAMOS DECIR QUE PARA RECORTAR ESPACIO AL HABER INCLUIDO NUEVOS COMENTARIOS
%
In contrast to the \textit{hetN} mutant, for the wild type and the \textit{patS} mutant skewness diminishes with time, as interval distributions become more symmetric.

%%%%%%%%%%%%% Figure  PatS mutant filament %%%%%%%%%%%%%%%%%%%%%%%
%%%%%%%%%%%%%%%%%%%%%%%%%%%%%%%%%%%%%%%%%%%%%%%%%%%%%

\subsection{Sequestration by PatS and HetN forms spatial gradients of HetR}

We have simulated the dynamics of gene expression in the model (see movies in the SI). 
%REALMENTE EN LAS FIGURAS NO SE PUEDE VER LO QUE SE DESCRIBE A CONTINUACION
%
Since PatS and HetN (or more accurately, the pentapeptide RGSGR represented by them in our theory) can diffuse along the filament, not surprisingly their profiles can form gradients \cite{kholodenko06,munoz2010}. At early times after nitrogen deprivation, PatS levels increase in the whole filament \cite{yoon98, yoon01,kumar10}. 
These levels remain high for some hours, but as some of these cells differentiate into heterocysts, system levels of PatS drop, and eventually high concentrations are observed only at new protoheterocysts that appear between formed heterocysts when filament growth moves them apart \cite{yoon01}.
%Fig.~\ref{Fig.PatS_WT}A.
%
The dynamics of HetN are somewhat reversed. At early times HetN is not observed \cite{bauer97} since there are no heterocysts
that produce it in the filament. At late times HetN cell concentrations form quasi-regular gradients with peaks at heterocysts  \cite{li02, callahan01}.

HetR concentration levels have been found to form interesting spatial profiles. For instance, expression of {\em hetR} has been shown to be correlated between neighbor cells before nitrogen deprivation \cite{corrales-guerrero2015}.
Since HetR does not diffuse, any gradient of its concentration would be produced by its interactions with the gradients of PatS and HetN. And this is precisely what occurs in our model: the expression of HetR shows higher values in cells far apart form heterocysts. The resulting HetR profile shows variability from cell to cell, just as observed experimentally \cite{risser09}.
Another prediction of this model is that the dynamics of the border cell is different from those in the middle, due to the accumulation of diffusing peptides at the end of the filament (see the Fig.~S4B for more details).

\subsection{Quasisynchronous cell division induces oscillatory variation of pattern properties}

\begin{figure}[t]
\begin{center}
\includegraphics[width=\linewidth]{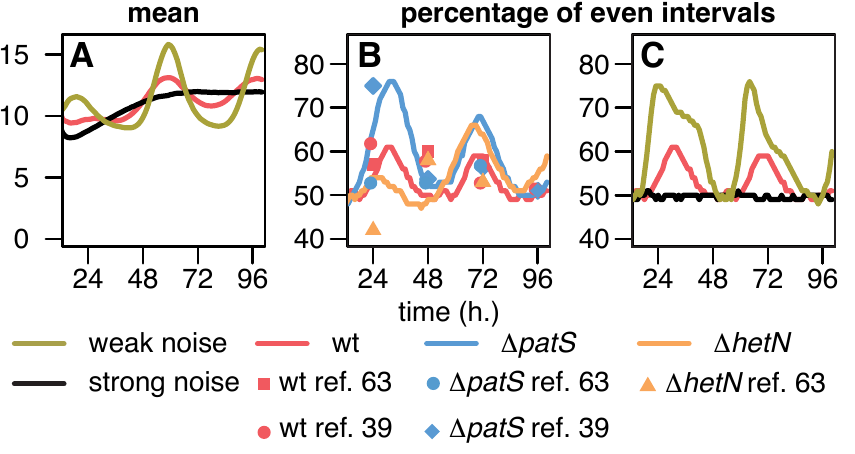}
\end{center}
%\caption{Quantitative dynamics of heterocyst patterning.
\caption{($A$) Effect of the noise in cell growth on the mean of the vegetative cells interval distribution.
($B$) Percentage of even intervals for the wild type, \textit{patS} and \textit{hetN} mutants in our model (lines) compared to experiments (symbols) as indicated. 
($C$) Effect of the noise in cell growth on the percentage of even intervals. For noise strength parameters, see Tables~S1 and S2.
%
%\highlightr{[Update ref. [6] to \cite{yoon01}, ref. [54] and ref. [55] are both \cite{corrales-guerrero2014plos}.]}
\label{Fig.Noise}}  
\end{figure}

The moments of the vegetative interval length distribution show an oscillatory behavior over time, especially the mean and the variance (Fig.~\ref{Fig.Results}).
The interpeak time in these oscillations is related to the average time for cell division.
When vegetative cells divide synchronously, the average distance (and variance) between heterocysts increases.
When a new heterocyst appears in the middle of an interval, it divides in two intervals roughly half in length.
If the cell cycle is roughly synchronous in the whole filament, this mechanism will produce a lengthening of the mean interval with each round of cell division, and a shortening with each round of differentiation.
In order to test this idea, we have made simulations with different levels of noise in cell growth dynamics (Fig.~\ref{Fig.Noise}A). Decreasing the noise, that is, synchronizing cell division, makes the oscillations of the mean interval distance more pronounced. Conversely, increasing the noise makes the oscillations disappear. The effects of the variation of the level of noise for other parameters are shown in the {\em Supporting Information} (Fig.~S6).

%Obviously this distance decreases when new heterocysts appear, until a new round of divisions occurs and it increases again. Clearly these oscillations are diminished when the intrinsic noise of the duplication is high as shown in Fig.~\ref{Fig.Noise}B where the mean distance between heterocysts is depicted for different values of the noise intensities associated to cell division. Thus, it is proved that asynchronous division due to different life cycle for the cell forming the initial filament or other factors induces a more regular pattern over time.
%
%This synchronous division is also associated to a large number of even intervals of vegetative cells between heterocysts. As noticed in Fig.~\ref{Fig.Noise}C when the growth noise is high cells do not tend to divide synchronously and the number of even intervals is reduced to a 50$\%$. 

%%%%%%%%%%%%% Figure HetN mutant filament %%%%%%%%%%%%%%%%%%%%%%%
%%%%%%%%%%%%%%%%%%%%%%%%%%%%%%%%%%%%%%%%%%%%%%%%%%%%%

%%%%%%%%%%%%% Figure Parity %%%%%%%%%%%%%%%%%%%%%%%
%%%%%%%%%%%%%%%%%%%%%%%%%%%%%%%%%%%%%%%%%%%%%%%%%%%%%

\subsection{Quasisynchronous cell division favors intervals of even length}

A characteristic property of the heterocyst pattern is the larger frequency of even-numbered vegetative cell intervals with respect to odd ones \cite{meeks02}. This is apparent from experimental histograms of interval length in the literature \cite{yoon98, yoon01, meeks02, khudyakov04, wu04, risser09}. 
Our theory reproduces this observation for wild type and mutants (Fig.~\ref{Fig.Noise}B). The fraction of even intervals also shows an oscillatory behavior over time, a sign that it is also caused by the synchronous character of cell division along the filament. To test this, we plot (Fig.~\ref{Fig.Noise}C) the percentage of even intervals for different levels of noise in cell growth.
For weak noise levels in cell growth, synchronous cell division induces a large fraction of even intervals. This fraction drops when interval lengths are large and several differentiation events happen almost simultaneously, randomizing interval parity until a new round of cell division occurs. In contrast, for high noise
rounds of cell division are not even well defined, and the percentage of even intervals remains always close to 50\%. 
%
%makes the oscillation of the mean interval distance more pronounced. Conversely, increasing the noise makes the oscillation disappear.
%
%but its peaks are anticorrelated with the peaks of the mean interval length. 

Our result supports previous suggestions \cite{khudyakov04} that the synchronous division of vegetative cells during the time that new heterocysts are formed is responsible for the higher fraction of even intervals, with no need of extra mechanisms.

\section{Discussion}

In this work we have shown that a model based on a positive feedback loop that promotes heterocyst differentiation, plus three diffusible inhibitors, an early one with source at protoheterocysts and two late ones with source at heterocysts, is enough to explain the formation and maintenance of heterocyst patterns up to quantitative detail.
As suggested by the biochemistry of the system \cite{feldmann11}, inhibition works through a {\em multimer cloud} mechanism \cite{schroeter12}, in which the formation of complexes between activators and inhibitors sequesters activator complexes and precludes them of realizing their activity. 
For clarity and concreteness we have only considered the main genetic regulators of the system, {\em hetR}, {\em patS}, {\em hetN}, as effective players for local activation, early long range inhibition, and late long range inhibition, respectively. 
A single late inhibitor would produce an overproduction of heterocysts in the {\em hetN} loss of function condition. This suggests the existence of, at least, a second late inhibitor, weaker and diffusing faster than HetN. A plausible candidate is the fixed nitrogen produced by heterocysts.
%\highlightr{The role of fixed nitrogen products is speculative: our theory strongly suggests the existence of a second late inhibitor, weaker than {\em hetN} and diffusing faster, but we cannot make a certain claim with regards to its identity.}
%However, note that other genes and metabolites contributing to these effects are taken into account in an effective way through the model parameters.
%as \textit{ntcA} for activation and nitrogen concentration for inhibition,
%
%We have shown that a single late inhibitor would invariable produce an overproduction of heterocysts much larger than observed in the {\em hetN} loss of function condition.
%This indicates that the effect of the second late inhibitor of differentiation is non-negligible, but also not enough by itself to explain the experimental patterns.

From a theoretical, more generic perspective, an interesting property of the model is that the main activator for heterocyst differentiation, HetR, is not diffusing.
The main mechanisms for the pattern formation and maintenance are
the positive feedback loop for HetR production and diffusion of the inhibitor molecules between cells.
The pattern shows a well defined intrinsic wavelength, robust in time, that does not significantly change due to small variations in model parameter values (see {\em Supporting Information} for a sensitivity analysis, Fig.~S7). This contrasts with the classical Turing continuum model for pattern formation in reactive-diffusive systems \cite{turing1952}, where the linear wavelength is proportional to the square root of the product of the activator and inhibitor diffusivities \cite{murray2001}. Another main difference with most Turing systems is that the conditions for pattern formation in Turing models need to be finely tuned, and patterns are generally non-robust in the sense that small variations in parameters may alter the observed wavelength, and even move the system out of the Turing regime. Actually, frequently in the Turing regime different patterns arise at the same point in parameter space, simply owing to slight variations in initial conditions. Obviously, this is not the case in heterocyst pattern formation, where a similar pattern is found independently of environmental and initial conditions.
%suggesting that indeed additional mechanisms control the pattern robustness.}

A key aspect of our work is bringing together regulatory aspects and filament growth through cell division. Without this last aspect, it is possible to create pattern forming models, but something as important as the maintenance of the pattern cannot be reproduced. Moreover, we show that it is not necessary to consider any crosstalk between the cell cycle and the regulation of heterocyst differentiation in order to reproduce experimental patterns.
This crosstalk, or some other mechanism, has been postulated as a part of a two stage model \cite{meeks02}, where this mechanism would form a coarse prepattern, from which in a second stage protoheterocysts are refined \cite{toyoshima10}.
Our model reproduces quantitatively experimental patterns with a purely stochastic pattern initiation.
Both small irregularities in the initial condition or the dynamic noise in the simulations are enough by themselves to trigger the emergence of the pattern.
Although a first stage is not needed to form a prepattern, we cannot exclude its existence. Evidence for it has been found in a different strain of cyanobacteria, {\em Nostoc punctiforme} strain ATCC 29133 \cite{risser2012}.

The mechanisms contemplated in our model are only the backbone of this patterning process. The phenotypes of other genes such as \textit{patA} and \textit{hetF} \cite{buikema01,liang92,wong01,risser08}
show that there is more to the system than local activation and long range inhibition. Future work should include modeling these finer mechanisms in order to be able to refine our understanding of how they work and discard hypothesis that do not pass the test of a theory-experiment comparison. Also, extension to different strains of cyanobacteria, and, importantly, use of live gene expression data as in \cite{corrales-guerrero2015} are other main avenues for future work.
A better understanding of biochemical kinetics and spatiotemporal aspects of gene expression and metabolite distribution would motivate the construction of models more experimentally constrained. This would allow to use these descriptions to obtain observable predictions and as guides to conduct novel experiments.\\

%La frecuencia de las oscilaciones en Maxsize/2/Growthrate

%El valor que uso en crecimiento lineal es de 0.05 (micrometros/h). Eso supone un tiempo de duplicación (mean generation time) de 2/0.05=40 h (Como en Allard et al que dicen que lo sacan e una pelicula de Golden) y por tanto un specific growth rate = ln(2)/40h= 0.017 (1/h). De hecho en las simulaciones es un poco más bajo (0.0165) debido a la aparición de heterocites. Creo que este es el orden de magnitud correcto según el paper de Pinzon. No tanto en el de Stacey (donde $\mu=2.3/24*k=0.1k$)

%% == end of paper:

%% Optional Materials and Methods Section
%% The Materials and Methods section header will be added automatically.

%% Enter any subheads and the Materials and Methods text below.
%\begin{materials}
% Materials text
%\end{materials}

%% Optional Appendix or Appendices
%% \appendix Appendix text...
%% or, for appendix with title, use square brackets:computational
%% \appendix[Appendix Title]

\section{Acknowledgments}
We would like to thank Joel Stavans, Fernando Falo and Jes\'us G\'omez-Garde\~nes for enlightening discussions, and Carlos Rasc\'on for critical reading of the manuscript.
We acknowledge funding from the Spanish Ministry of Economy and Competitiveness (MINECO) through Grant PHYSDEV (No. FIS2012-32349) and the Ram\'on y Cajal program (S. A.).
%\end{acknowledgments}

\setcounter{figure}{0}

\makeatletter 
\renewcommand{\thefigure}{S\@arabic\c@figure}
\makeatother

\makeatletter 
\renewcommand{\thetable}{S\@arabic\c@table}
\makeatother

\section{Supporting Information}

\section{Movie captions}

\noindent Supplementary movies available at

\noindent \url{http://bit.ly/299J9IU}

\noindent {\bf Movie S1}: Temporal evolution of gene expression in a filament for wild type. Time counts hours after nitrogen deprivation.  Heterocyst cells have a yellow membrane. 
The intensity of the green, blue, red, and cyan colors show the level of HetR, PatS, HetN and fixed nitrogen products concentrations respectively.
When a filament is too long to fit in the width of the movie, it is continued in a row below. The last cell on the right of a row is a neighbor of the first cell on the left of the row immediately below.\\

\noindent {\bf Movie S2}: Temporal evolution of gene expression in the filament for {\em patS} loss of function. Time counts hours after nitrogen deprivation. Heterocyst cells have a yellow membrane. 
The intensity of the green, red, and cyan colors show the level of HetR, HetN and fixed nitrogen products concentrations respectively.
In the second line, cells are black because they do not express PatS.
When a filament is too long to fit in the width of the movie, it is continued in a row below. The last cell on the right of a row is a neighbor of the first cell on the left of the row immediately below.\\

\noindent {\bf Movie S3}: Temporal evolution of gene expression in a filament for {\em hetN} loss of function. Time counts hours after nitrogen deprivation. Heterocyst cells have a yellow membrane. 
The intensity of the green, blue, and cyan colors show the level of HetR, PatS, and fixed nitrogen products concentrations respectively.
In the third line, cells are black because they do not express HetN.
When a filament is too long to fit in the width of the movie, it is continued in a row below. The last cell on the right of a row is a neighbor of the first cell on the left of the row immediately below.\\

\noindent {\bf Movie S4}: Temporal evolution of gene expression in a filament for {\em hetN} and fixed nitrogen loss of function. Time counts hours after nitrogen deprivation. Heterocyst cells have a yellow membrane. 
The intensity of the green and blue colors show the level of HetR and PatS concentrations respectively.
In the third and fourth lines, cells are black because they do not produce HetN or fixed nitrogen products.
When a filament is too long to fit in the width of the movie, it is continued in a row below. The last cell on the right of a row is a neighbor of the first cell on the left of the row immediately below.\\

\noindent {\bf Movie S5}: Temporal evolution of gene expression in a filament for fixed nitrogen loss of function. Time counts hours after nitrogen deprivation. Heterocyst cells have a yellow membrane. 
The intensity of the green, blue, and red colors show the level of HetR, PatS, and HetN concentrations respectively.
In the fourth line, cells are black because in this condition fixed nitrogen products have no function in heterocyst differentiation.
When a filament is too long to fit in the width of the movie, it is continued in a row below. The last cell on the right of a row is a neighbor of the first cell on the left of the row immediately below.\\

\section{Model construction} \label{modcon}

We have constructed a model consisting of four variables per cell representing the concentration in the cell of molecular species responsible for the mechanisms of local autoactivation, early long range inhibition and two types, strong and weak, of late long range inhibition. We identify the molecules responsible for these mechanisms with HetR, PatS, HetN and fixed nitrogen products, and base the reaction scheme between the genes (Fig.~1 of the Main Text)
on the biochemistry discussed in [41].
%\cite{feldmann11}
%The weak late inhibitor is introduced in a phenomenological way, and its identification with fixed nitrogen products is speculative. 
The identification of the weak late inhibitor with fixed nitrogen products is speculative.
However, they are a strong candidate to play this role and in the following we will refer to them when discussing this mechanism. For simplicity we do not describe mRNA concentrations and, for each cell $i$, have four variables: $h_i$ for the concentration of HetR monomers, $p_i$ for the concentration of PatS monomers, $n_i$ for the concentration of HetN monomers, and $f_i$ for the concentration of fixed nitrogen products.
% Under a {\it hetN} loss of function condition, $n_i$ may represent in an phenomenological way the effect of fixed nitrogen products, we shall discuss this later. 

In our model we distinguish between vegetative cells and heterocysts. In vegetative cells, HetN is not produced. HetR is produced with a small constitutive rate [34].%\cite{rajagopalan10}.
Its monomers can bind into dimers [42,50,51].
%\cite{huang04,kim2011,kim2013}.
These HetR dimers can bind the \textit{hetR} and \textit{patS} promoter regions and activate production of HetR and PatS.
If the concentration of HetR builds up above a threshold value without ever falling below a certain minimum value, the vegetative cell differentiates and turns into a heterocyst.
Since heterocysts have already differentiated and HetR cannot diffuse between cells in our model, modeling HetR concentration in them is superfluous. However, since experimentally HetR concentration in heterocysts is high, we model it with a constitutive term. Heterocysts do not produce PatS, but HetN and fixed nitrogen products are produced with a constitutive rate.

The RGSGR pentapeptide, a product of PatS, can bind HetR with a 1:1 stoichiometry, and RGSGR-tagged HetR molecules can no longer bind DNA [41].
%\cite{feldmann11}.
We assume the concentration of RGSGR derived from PatS is proportional to $p_i$.
HetN also contains a RGSGR motif [45],
%\cite{higa12},
so as with PatS, we assume the concentration of RGSGR derived from HetN is proportional to $n_i$.
We allow the RGSGR pentapeptide to diffuse between neighbor cells, irrespectively of their vegetative or heterocyst identity. Because of this, we can find non-zero levels of $p_i$ in heterocysts or non-zero levels of $n_i$ in vegetative cells.
Finally, we assume that HetR dimers tagged with one or two RGSGR molecules can no longer bind DNA [41].
%\cite{feldmann11}.
%
Hence, inhibition of HetR activity by PatS or HetN would be due to sequestration or titration of HetR dimers.
This is similar to the {\em cloud of dimers} mechanism that appears in other biological processes [69].
%\cite{schroeter12}.
%

Fixed nitrogen products are assumed to diffuse between cells much faster than the RGSGR pentapeptide and be weaker inhibitors than HetN and PatS. 

\section{Mathematical formalism}

The model description above can be used to write a set of chemical reactions for the following species: HetR monomers, HetR dimers (HetR:HetR), PatS monomers, HetN monomers, fixed nitrogen products, and the complexes HetR:HetR:PatS, HetR:HetR:HetN, HetR:HetR:Pats:PatS, HetR:HetR:HetN:HetN and HetR:HetR:PatS:HetN. PatS and HetN in the complexes actually represent RGSGR molecules.
Only monomers can be produced by transcription and translation (bundled into a single process). We consider linear degradation kinetics for both monomers and multimers.
Formation of a complex between HetR:HetR and the promoter regions of {\em hetR} and {\em patS} in vegetative cells is also considered. The formation of these complexes increases the rate of HetR and PatS production, respectively. The effect of fixed nitrogen products is included in a phenomenological way as an inhibition in the regulatory function controlling the production rate of HetR and PatS.
Assuming mass-action kinetics we can write differential equations for the time evolution of each species.
Making the usual assumption that all the multimer reactions (binding and unbinding of monomers) are much faster than degradation reactions [72],
%\cite{spinner02},
we can make an adiabatic approximation and eliminate all multimer concentrations from the description. The resulting kinetics for monomer concentrations are
\begin{align}
\frac{d h_i(t)}{dt}=&\, b_h +  a_{h} g(h_i,p_i,n_i,f_i) - \alpha_h h_i(t)\left[1+2\mu h_i(t)\right],  \label{Eq.HetR} \\
\frac{d p_i(t)}{dt}=& \, b_p+ [1- \delta_{\text{hc},i}] a_p g(h_i,p_i,n_i,f_i) -\alpha_p p_i(t)+\nonumber \\ &d_p \left[p_{i+1}(t)-2p_i(t)+p_{i-1}(t)\right], \label{Eq.PatS} \\
\frac{d n_i(t)}{dt}= & \, b_n + \delta_{\text{hc},i} a_n -\alpha_n n_i(t) +\nonumber \\ & d_n \left[n_{i+1}(t)-2n_i(t)+n_{i-1}(t)\right], \label{Eq.HetN} \\
\frac{d f_i(t)}{dt}= & \, b_f + \delta_{\text{hc},i} a_f -\alpha_f f_i(t) +\nonumber \\ & d_f \left[f_{i+1}(t)-2f_i(t)+f_{i-1}(t)\right], \label{Eq.fixN}
\end{align}
with
{\footnotesize
\begin{align}
&g(h_i,p_i,n_i,f_i)= \nonumber \\ & \frac{ \left[\frac{h_i(t)}{k}\right]^2}{1 + \left[\frac{h_i(t)}{k}\right]^2 + \frac{p_i(t)}{K_d} + \frac{n_i(t)}{K_d} + \left[\frac{p_i(t)}{K_d}\right]^2 + \left[\frac{n_i(t)}{K_d}\right]^2 + \frac{p_i(t) n_i(t)}{K_d^2} + \frac{f_i(t)}{K_f}}.  \label{function}
\end{align}}
The subindex $i$ means that the variable is referred to cell $i$ in the filament, its neighbor cells are $i-1$ and $i+1$ (cells at the extremes of the filament only have one neighbor), $t$ denotes time, $h_i$ is the HetR concentration in cell $i$, $p_i$ the PatS concentration, $n_i$ the $HetN$ concentration, and $f_i$ the fixed nitrogen products concentration. In a parameter, the subindex $h$ means that it is related to variable $h_i$, $p$ to $p_i$, $n$ to $n_i$ and $f$ to $f_i$.
$b_h$, $b_p$, $b_n$ and $b_f$ are basal or constitutive production rates; we have used $b_p=b_n=b_f=0$, but checked that the model is robust to finite values of these parameters. $a_h$, $a_p$,  $a_n$ and $a_f$ are the maximum regulated production rates.
$\alpha_h$, $\alpha_p$, $\alpha_n$ and $\alpha_f$ are linear degradation rates. 
$d_p$, $d_n$ and $d_f$ are intercellular diffusion rates. We take the concentration of each species to be homogeneous inside a cell.
The nonlinear degradation term in the equation for $h_i$ appears as the effect of dimer-mediated degradation of monomers [69].
%\cite{schroeter12}.
The parameter $\mu$ is related to the rates $k_b$ of HetR monomers binding to form HetR dimers, $k_u$ of HetR dimers unbinding into HetR monomers, $\alpha_d$ of dimer degradation and $\alpha_h$ of monomer degradation as
\begin{equation}
\mu = \frac{k_b}{\alpha_h}\left(1-\frac{k_u}{k_u + \alpha_d}  \right).
\end{equation}
The factor $ \delta_{\text{hc},i}$ specifies if the cell is vegetative or a heterocyst: its value is $1$ if cell $i$ is a heterocyst, and $0$ if it is vegetative. Since heterocysts cannot divide, initially all cells are vegetative and have $\delta_{\text{hc},i}=0$; it changes to $\delta_{\text{hc},i}=1$ if at any point the cell differentiates into a heterocyst.
Regulation of the production of $h_i$ and $p_i$ is done through the function $g(h_i,p_i,n_i,f_i)$ specified in Eq.~(\ref{function}). $k$ is the equilibrium constant (or dissociation constant) of the binding-unbinding reaction between the HetR dimer and the {\em hetR} and {\em patS} promoters; for simplicity, we take both these constants to have the same value.  
$K_d$ is the equilibrium constant for the binding-unbinding reaction between HetR and the RGSGR pentapeptide. Its value has been measured experimentally for PatS-derived RGSGR: $K_d=227\pm 23$ nM [41].
%\cite{feldmann11}.
We assume the same value for HetN-derived RGSGR.
The different terms with $K_d$ in the denominator of Eq.~(\ref{function}) come from the titration of HetR dimers by a single PatS-derived RGSGR molecule, a single HetN-derived RGSGR molecule, two PatS-derived RGSGR molecules, two HetN-derived RGSGR molecules, and one PatS-derived plus a HetN-derived RGSGR molecules, respectively. The function could have been approximated by a simpler one where only the $p_i^2$ and $n_i^2$ terms related to $K_d$ would be taken into account: this would have not changed the dynamics of the model in any relevant way. For the effective constant $K_f$ controlling the strength of HetR and PatS repression by fixed nitrogen products, we have chosen to use the same value as for $K_d$.
Taking $K_f=K_d$
may be unrealistic. However, the strength of the effect of the fixed nitrogen products is given by both $K_f$ and by its production rate $a_f$. To make this inhibition effect weaker than the effect of HetN, we have set the production rate of the fixed nitrogen products to be much lower than the production rate of HetN, $a_f \ll a_n$.

% only real parameter here is the relation between $a_f$ and $K_d$, so we have tuned the effect of fixed nitrogen setting its production rate to be much lower than the production rate of HetN, $a_f \ll a_n$.

\section{Model implementation}

We have implemented a code in an object-oriented platform to model the differentiation and growth of vegetative and heterocyst cells. In this programming environment each cell is described by an agent which has its own variables representing cell size and the current concentration of the species considered, HetR, PatS, HetN, and fixed nitrogen products. These agents are ordered in a structure representing the filament. Because there can be a net flux of protein concentration between cells, the evolution equations for each agent are coupled with the equations for its adjacent neighbors. Cells at the extremes of the filament only have one neighbor.
The evolution of the protein and fixed nitrogen products concentrations for each cell is based on the noisy version of Eqs.~(\ref{Eq.HetR})-(\ref{Eq.fixN}). 
The stochastic nature of gene expression
%due to the molecular low copy number
has been considered in our model by extending the equations derived above to the Langevin dynamics in the It\^o interpretation [73].
%]\cite{gillespie00,bialek01,gardiner04}.
Since the evolution for a given concentration at cell $i$, $x_i$, consists on a sum of production (synthesis) terms, $P_i$, minus the sum of degradation terms, $D_i$, a stochastic term of the form $\eta_{xi}(t)\sqrt{P_{xi}+|D_{xi}|}$  was included in Eqs.~(\ref{Eq.HetR})-(\ref{Eq.HetN}). Here $\eta_{xi}(t)$ is an uncorrelated Gaussian white noise with zero mean and variance $\langle \eta_{xi}(t) \eta_{xi'}(t') \rangle = \delta_{ii'} \delta(t-t')/V $, where $V$ is an nondimensional effective volume setting noise strength in gene expression [73].
%\cite{frigola12}.

Vegetative cell growth was modeled by a stochastic differential equation for each agent:
\begin{equation} \label{growth}
\frac{d s_i(t)}{d t}= \rho + \eta_{si}(t),
\end{equation} 
where $s_i$ is the size of cell $i$, $\rho$ is a constant growth rate, and $\eta_{si}(t)$ is an 
uncorrelated Gaussian white noise
with zero mean and variance $\langle \eta_{si}(t) \eta_{si'}(t') \rangle = G \delta_{ii'} \delta(t-t') $,
which models the intrinsic fluctuations in the growth process. 
Starting from an initial size, each cell evolves following Eq.~(\ref{growth}) up to a maximum size $\eta_{di}$, which is a noisy value drawn from a Gaussian distribution for each cell.
When this size is reached the vegetative cell divides, producing two new vegetative cells with half of its current size and identical protein concentrations.
Heterocysts grow following the same dynamics, but once reached their maximum size, $\eta_{di}$, they do not divide and stop growing.

To differentiate into a heterocyst, a vegetative cell has to build up a certain level of HetR. This has been implemented integrating over time for each vegetative cell the value of HetR concentration, once the value of $h_i$ is above a threshold $\eta_{rs}$. This threshold is cell specific, being drawn from a Gaussian distribution. If at any point the value of $h_i$ drops below $\eta_{rs}$, the integral is reset to zero. Otherwise, if the integral ever reaches a value $\eta_{ci}$, also a cell-specific Gaussian distributed parameter, the vegetative cell differentiates into a heterocyst and $\delta_{\text{hc},i}$ changes from $0$ to $1$. 
An alternative way to model the differentiation decision is to consider that if $h_i$ is above a given threshold for a certain window of continuous time, that particular vegetative cell will transform into a heterocyst. We checked this procedure, using the same threshold $\eta_{rs}$ listed in Table~S2
%\ref{tab.noise}
and a time window of 8 hours, which has produced results indistinguishable from the method based on integrating the concentration over time.
%Our choice of differentiation method came from technical reasons regarding simplicity of the simulation procedure.
%
%However, as we will see in the next section, this simple description based on Eqs.\ (\ref{eq.min3xxxx})-(\ref{function}) will allow us to reproduce some of the main features observed in these systems when the available nitrogen is step-down and understand many processes in heterocyst differentiation. In the next section we will discuss some main results and compare with previous experimental works.
%

Numerical integration of the equations has been made using the Euler-Maruyama approximation for stochastic differential equations.
%\cite{kloeden92}.
Each simulation starts with 150 cells of size 3.3 $\mu$m, and initial concentrations at each cell $p_i(0)=n_i(0)=0$ nM and a noisy value drawn from a Gaussian distribution with mean $9.4$ nM and standard deviation 0.2 nM for $h_i(0)$.
No-flux boundary conditions were employed.
The beginning of the simulation represents the moment of nitrogen deprivation. The statistical results in this work come from an average over 192 different realizations.

To simulate loss of function conditions, we have made the relevant parameters equal to zero. For instance, 
for the {\em patS} loss of function condition, we used $a_p=0$ nM/h, ensuring no PatS production.
Overexpression is simulated setting the corresponding basal production rate equal to the maximum regulated production rate.

To calculate the mean, variance and skewness of vegetative cell intervals (Fig.~4 of the main text), we have not considered intervals of length 0, frequent in a Mch phenotype. The reason to do this is to allow comparison with experimental references that do not report this piece of information.

\section{Parameter estimation}

Some parameters were obtained from literature. The affinity of RGSGR to HetR, $K_d$, was taken directly from [41].
%\cite{feldmann11}. 
%
We consider that cells that have just divided are about 2 $\mu$m long and roughly double their size before division. With these numbers, we have considered a cellular growth rate of $\rho=0.05$ $\mu$m/h, which produces a mean generation time of 40 h, in agreement with the data from [56]
%\cite{khudyakov04}
analyzed in [13].
%\cite{allard07}.
%
All the other parameters were chosen with the aim to quantitatively reproduce the observations of the heterocyst pattern dynamics for the wild-type and mutant phenotypes. A preliminary linear stability analysis of a continuous version of Eqs.~(\ref{Eq.HetR})-(\ref{function}) was performed in order to establish a valid range of parameters in which the homogeneous steady state of the system is unstable and a spatial pattern in the protein concentration levels appears. 
%
%Analytical predictions for this continuous version were validated by numerical integration using a fourth-order Runge-Kutta method for time and a standard centered finite difference scheme for spatial derivatives with periodic boundary conditions.
%
Parameter values used in the wild-type simulation are given in Table~S1.%\ref{tab.parameters}.
%bFor the {\em patS} loss of function condition, we used $a_p=0% nM/h, ensuring no PatS production. The {\em hetN} loss of function condition is more complicated, see below.

Together with the dynamic noise in gene expression and cell growth, some parameters have cell specific values obtained from a Gaussian distribution using the Box-Muller algorithm.
%\cite{box1958note}.
The mean and standard deviation of these parameters are given in Table~S2.%\ref{tab.noise}.

%\highlightr{El valor que uso en crecimiento lineal es de 0.05 (micrometros/h). Eso supone un tiempo de duplicacion (mean generation time) de 2/0.05=40 h (Como en Allard et al que dicen que lo sacan e una pelicula de Golden) y por tanto un specific growth rate = ln(2)/40h= 0.017 (1/h). De hecho en las simulaciones es un poco mas bajo (0.0165) debido a la aparici0n de heterocites. Creo que este es el orden de magnitud correcto segun el paper de Pinzon. No tanto en el de Stacey (donde $\mu=2.3/24*k=0.1k$).
%
%The averaged specific growth rate in the simulations for the considered parameters after four days was 0.0165 h$^{-1}$.

\section{Temporal evolution of gene expression in the filaments}

\begin{figure*}[!htpb]
\centering
\includegraphics[width=0.8\textwidth]{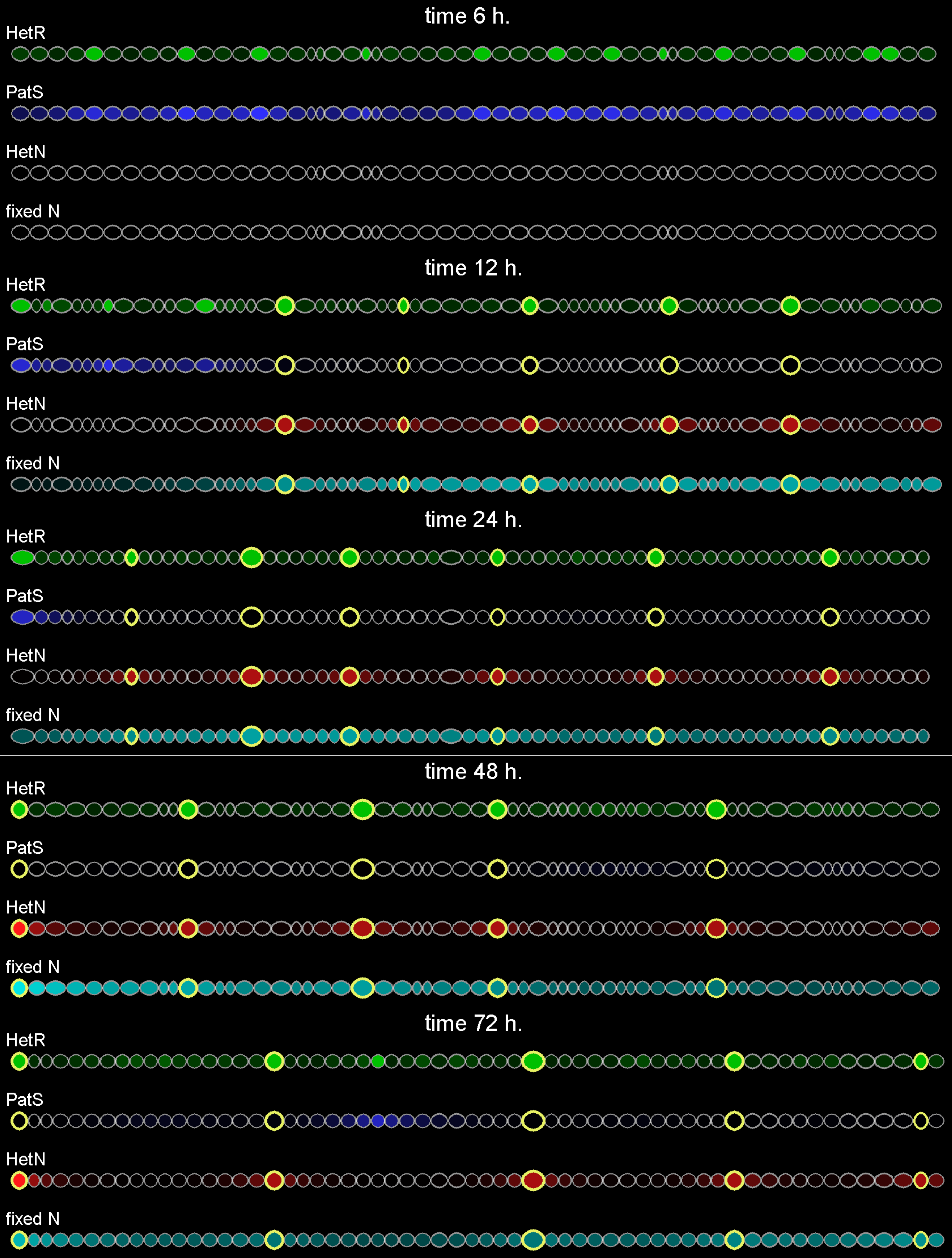}
\caption{Temporal evolution of gene expression in a filament for wild type, depicted at 6, 12, 24, 48, and 72 hours after nitrogen deprivation.  Heterocyst cells have a yellow membrane. 
The intensity of the green, blue, red, and cyan colors show the level of HetR, PatS, HetN and fixed nitrogen products concentrations respectively.
For convenience, in order to occupy a single line only the initial part of the filament is shown. See also Movie S1.}
\label{sfig.expression_wt}
\end{figure*}

\begin{figure*}[!htpb]
\centering
\includegraphics[width=0.8\textwidth]{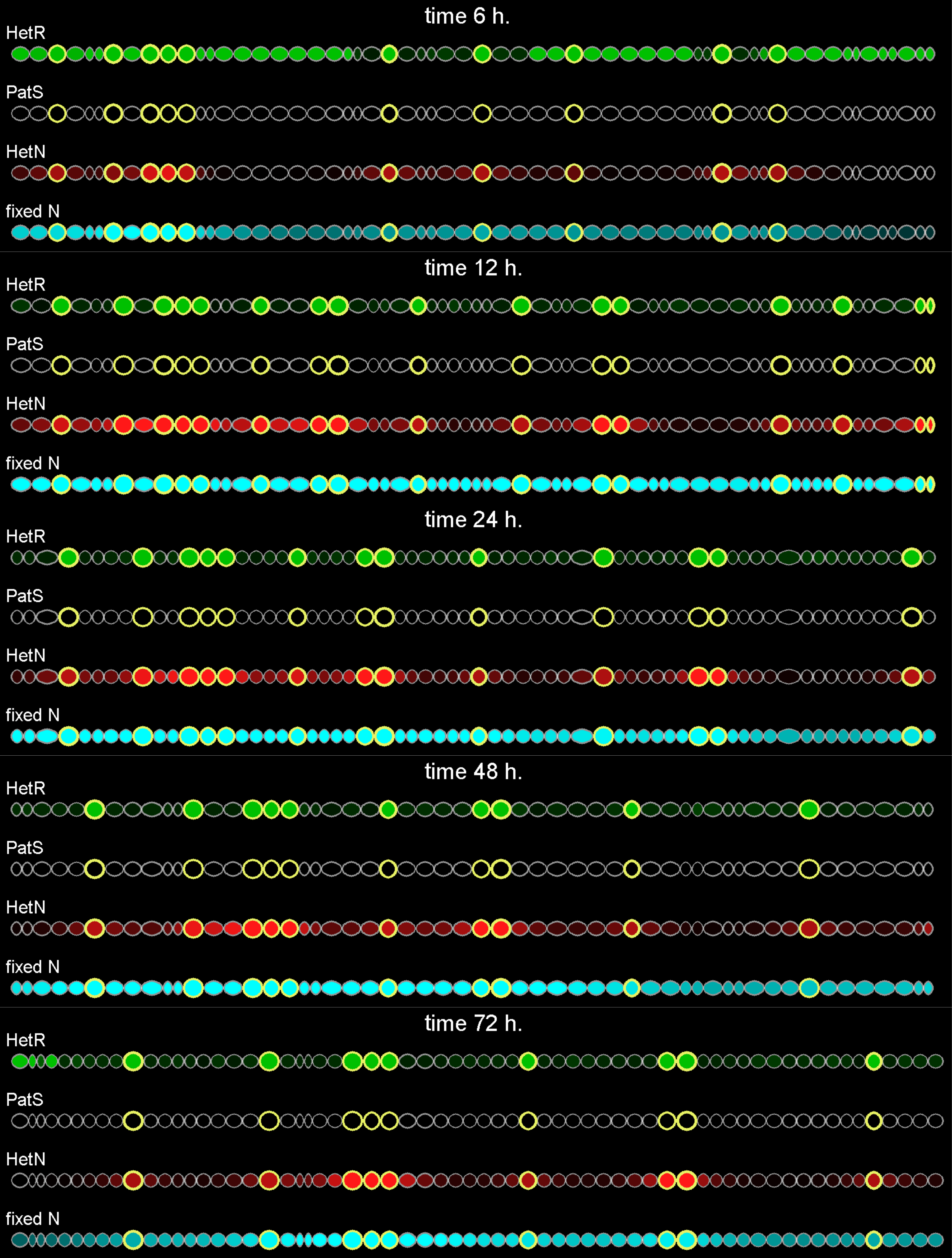}
\caption{Temporal evolution of gene expression in a filament for {\em patS} loss of function.
Details as in Fig.~\ref{sfig.expression_wt}.  See also Movie S2.}
\label{sfig.expression_pats}
\end{figure*}

\begin{figure*}[!htpb]
\centering
\includegraphics[width=0.8\textwidth]{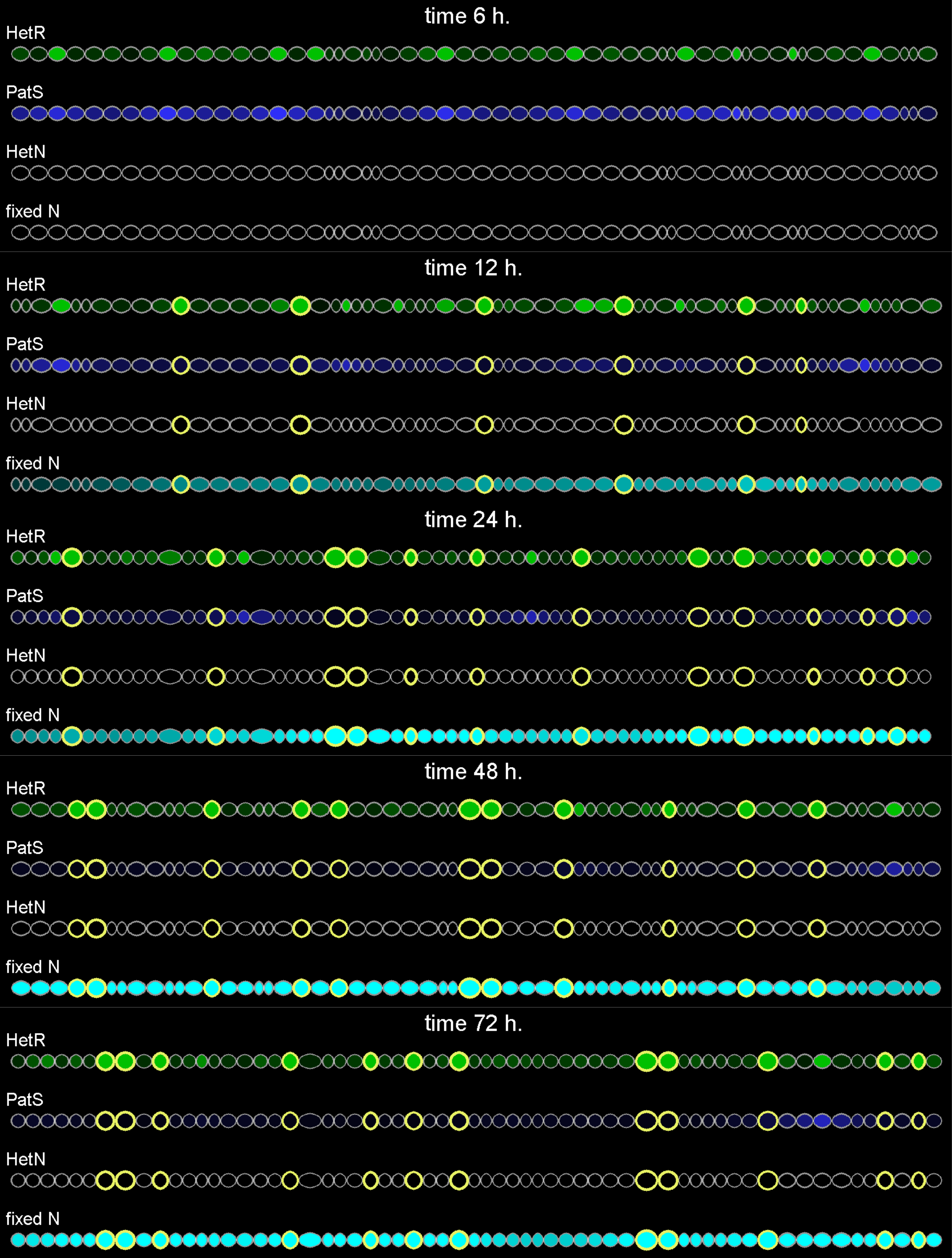}
\caption{Temporal evolution of gene expression in a filament for {\em hetN} loss of function.
Details as in Fig.~\ref{sfig.expression_wt}.  See also Movie S3.}
\label{sfig.expression_hetn}
\end{figure*}

Figs.~\ref{sfig.expression_wt}-\ref{sfig.expression_hetn} and \ref{sfig.expression_fixn} show the expression of all four species considered in the model for different conditions at 6, 12, 24, 48, and 72 hours after nitrogen deprivation. Some of the main features of the temporal evolution of these concentrations for each phenotype have been already discussed in the main text. For instance, as already stressed there, high levels of PatS are found for the wild type at short times (Fig.~\ref{sfig.expression_wt}). However, PatS concentrations decrease with time and, eventually, large gradients of PatS are found only around protoheterocysts at long times. The dynamics of HetN is different. Gradients of HetN are only observed next to heterocysts at late times. Regarding to the differences between wild type and mutants, HetR levels are larger in the {\em patS} mutant than in the wild type and {\em hetN} mutant at short times, coming back to low levels after a few hours (Fig.~\ref{sfig.expression_pats}). This agrees with experimental observations [23,39].
%\cite{yoon98,yoon01}.
Another main difference between the wild-type and the mutant phenotypes is that the appearance of contiguous heterocysts is more common in the mutants. However, in contrast to the early contiguous heterocysts in the {\em patS} mutant, in the {\em hetN} mutant they form at later times (Fig.~\ref{sfig.expression_hetn}). Thus, if we inspect a {\em hetN} mutant a long time after nitrogen deprivation, a large amount of contiguous heterocysts can be found.

\section{Differentiation of border cells}

\begin{figure*}[!htpb]
\centering
\includegraphics[width=1.5\columnwidth]{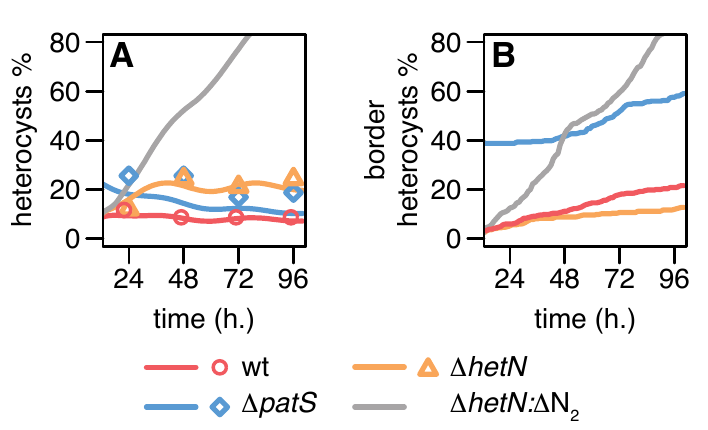}
\caption{Dynamics of the percentage of heterocysts.
($A$) Time evolution of the percentage of heterocyst cells in our model (lines) and experimental results in [47] 
%\cite{borthakur05}
(symbols). Red and circles for wild type, blue and diamonds for {\em patS} loss of function, orange and triangles for {\em hetN} loss of function, gray for no late long range inhibitors.
($B$) Time evolution of the percentage of heterocysts in border cells in our model (dashed lines). } \label{fig.het_percent}
\end{figure*}

\begin{figure*}[!htpb]
\centering
\includegraphics[width=0.8\textwidth]{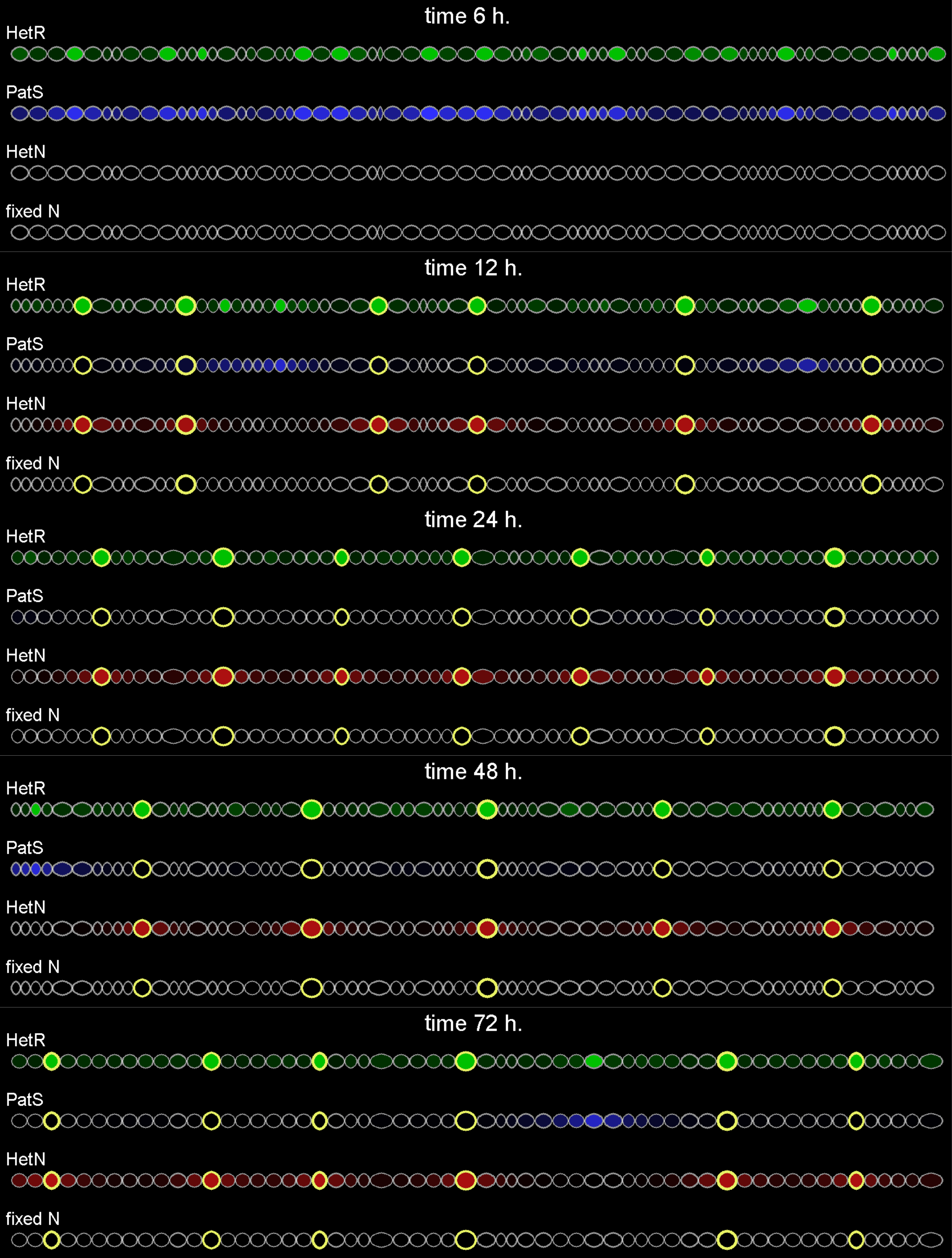}
\caption{Temporal evolution of gene expression in a filament for fixed nitrogen loss of function.
Details as in Fig.~\ref{sfig.expression_wt}.  See also Movie S5.}
\label{sfig.expression_fixn}
\end{figure*}

The cells at the border of a filament are an interesting study case, because the possible accumulation of diffusing peptides may render their dynamics different from those of the cells in the middle of a filament. An extreme case is the {\em patA} mutant, in which mainly the border cells of the filament differentiate into heterocysts [33,35,37].
%\cite{buikema01,liang92,risser08}.
In Fig.~\ref{fig.het_percent}B we plot the frequency of heterocyst formation both in the whole filament and considering only the two border cells, for wild type, {\em patS} loss of function, {\em HetN} loss of function, and no late long range inhibitors.
%As noted,
In all cases the probability of a border cell being a heterocyst grows over time, because once a border cell differentiates it will remain a heterocyst forever.

%En Orozco 2006 se dice ``A hypothesis that accounts for the unusual patA phenotype and the apparent epistasis of patS at 24 h postinduction predicts a role for patA in attenuation of the patS signal responsible for suppression of differentiation....'' Esto es interesante, porque justificaría en cierta forma que aumente los heterocistes en los bordes. De hecho está aquí explicado. Creo que tenemos otro paper ;)

Regarding the wild type, initially border cells have a lower propensity to differentiate than other cells. This is due to the fact that no-flux boundary conditions were employed in our model. Therefore, PatS does not diffuse at the borders and accumulates at them, which has an inhibitory effect on differentiation. However, once the first round of differentiation has occurred, the main role inhibiting differentiation is played by HetN, which is only produced in heterocysts and diffuses to their neighbor cells. Now, a border cell will be the one in its vegetative interval that is further away from the first heterocyst, so as cell division goes on moving that heterocyst further away from it, it will be the cell with higher chances to differentiate. For this reason the probability of a border cell to become a heterocyst becomes larger than in the interior of the filament.

The initial situation in a {\em hetN} mutant is very similar to the wild type. However, after the first round of differentiation, the inhibitory effect on heterocyst formation occurs only through PatS and fixed nitrogen products. The latter diffuses very fast and has an almost homogeneous distribution along the filament, so it does not significantly increase the border cell's advantage to differentiate over the rest of the cells. Thus, the initial situation for the cells at the end of the filaments of having a smaller probability of transforming into a heterocyst is maintained due to PatS accumulation. For this reason, in {\em hetN} mutants border heterocysts are less frequent than interior heterocysts.

The situation with no late inhibitors of any kind at early times is again similar to wild type, but as time goes by the probability to differentiate at the borders grows dramatically, just as in cells in the interior of the filament (Fig.~\ref{fig.het_percent}A).

In {\em patS} mutants, due to the lack of PatS accumulation at the borders, right after nitrogen deprivation there is not any inhibitor that stops a border cell from differentiating. Moreover, after the formation of the first generation of heterocysts, as in the wild type, the effect produced by HetN favors even higher probability of differentiation at the borders. Fig.~\ref{fig.het_percent}B shows that our simulations predict a very high frequency of border heterocysts in {\em patS} mutants. 

In summary, PatS and HetN have opposite effects on the probability of a border cell to become a heterocyst. Due to its accumulation at the borders in this theoretical model, PatS diminishes this probability.For HetN, since cells at the ends of the filament are the ones further away from the first heterocyst producing HetN, their probability to become heterocysts is larger than for the rest of the cells. It would be very interesting to check these predictions with experimental observations.

\section{Noise effects}

In the simulations shown in Fig.~7A and C of the main text three different pairs of values of the intrinsic noise in cell growth $G$ and the standard deviation of the maximum cell size $\eta_{di}$ were used in order to reproduce weak, medium, and strong noise conditions in cellular growth. The values of the parameters or their standard deviations employed for the weak and strong conditions in these simulations, and for the conditions discussed throughout this section, are indicated in parentheses in Table~S1 and Table~S2.%\ref{tab.parameters} and Table~\ref{tab.noise}.

\begin{figure*}[!t]
\centering
\includegraphics[width=1.5\columnwidth]{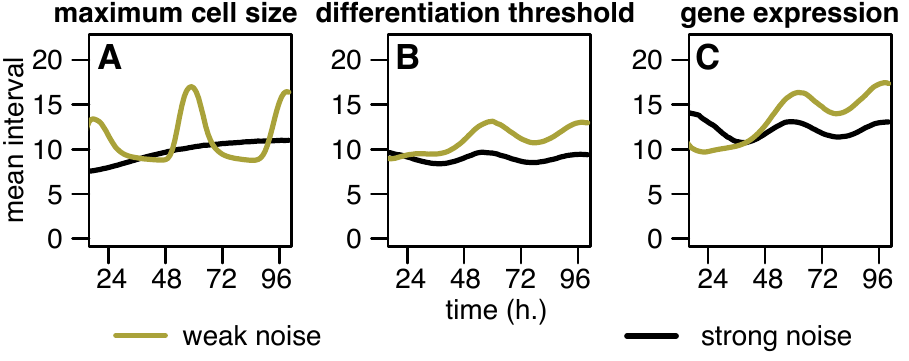}
\caption{Effect of noise on the mean of the vegetative cells interval distribution.
Noise in: ($A$) $\eta_{di}$, the maximum cell size for division,
($B$)  $\eta_{rs}$ and  $\eta_{ci}$, the maximum cumulative level and minimum reset level for differentiation, and ($C$) gene expression, given by the value of $V^{-1}$. Weak and strong stand for conditions with less and more variability than our wild-type parameters, respectively. Parameter values for these conditions are given in Tables~S1 and S2.}%\ref{tab.parameters} and \ref{tab.noise}.}
 \label{fig.mean_noise}
\end{figure*}

We have also studied the effect of variability on $G$ and $\eta_{di}$ independently. Weak and strong noise in cell growth and maximum cell size have a similar effect as that observed in Fig.~7A of the main text. Small values of these noises induce large oscillations in the mean interval distance between heterocysts. Large values decrease these oscillations, (see, for example, Fig.~\ref{fig.mean_noise}A for small and large deviations for the maximum cell size, $\eta_{di}$).

Increasing variation in the threshold, parameters $\eta_{rs}$ and $\eta_{ci}$ employed to decide when a cell differentiates, produces a small decrease of the interval mean (Fig.~\ref{fig.mean_noise}B). With a less homogeneous threshold, it is more probable that protoheterocysts close to mature heterocysts have small values of this threshold and differentiate easily. More regular and longer intervals are found with a weak variation in this parameter.

A similar effect as for the variation in $\eta_{rs}$ is found for the stochasticity in gene expression. Strong noise (small values of $V$) produces a decrease in the mean interval lengthg (Fig.~\ref{fig.mean_noise}C).
%
%For large values of $V$ a more regular pattern with large periodicity and small values of HetR concentrations appears.
%
That is because large noise in gene expression dynamics makes it easier for protoheterocysts to reach high values of HetR concentration and differentiate, which leads to shorter vegetative intervals.
Although there is no experimental data to make comparisons, a more detailed study of noise effects could represent an interesting direction for future work.

\section{Sensitivity analysis and robustness}

%\subsection{The pattern is robust with respect to changes in parameter values and initial conditions} 

We have performed a sensitivity analysis of the model [75] %\cite{savageau71}
and find that the mean distance between heterocysts and other pattern features are robust in time and respect to changes in model parameter values. The model is more sensitive to alterations in those parameters related to {\em hetR} production and degradation rates, and its affinity to promoters, stressing the importance of the HetR local positive feedback, and confirming that {\em hetR} acts as the master regulator of the process. In any case, for the considered parameter ranges the system shows a stable heterocyst pattern with a well defined wavelength for any initial and boundary conditions.

%In order to gain insight on how small changes in parameter values affect the pattern and its robustness,
%we performed a sensitivity analysis [79,80].
%\cite{savageau71Nature,savageau71Archives}.
We define the sensitivity $S_{YX}$ of the observable $Y$ with respect to changes in the parameter $X$ as:
\begin{equation}
 S_{Y X}=\frac{\partial \log Y}{\partial \log X}.
\end{equation}
This quantity is evaluated at some point in the parameter space. The logarithmic derivative defined above can be written as:
\begin{equation}
  S_{YX}=\frac{dY/Y}{dX/X}. \label{Sderivative}
\end{equation}
To compute the sensitivity for small changes in the parameter $X$, we approximate equation (\ref{Sderivative}) by the ratio of relative changes as:
\begin{equation}
  S_{YX}=\frac{\Delta Y/Y}{\Delta X/X}.
\end{equation}
We determine the sensitivity of the model to changes in a given parameter $X$ by evaluating the selected observable $Y$ at two points, the focal point $X_0$ and $X= X_0 + \Delta X$. The sensitivity, $S_{YX}$, thus defined is the ratio between the percentage change in the variable $Y$ and the percentage change in the parameter $X$.

\begin{figure*}[!t]
\centering
\includegraphics[width=1.9\columnwidth]{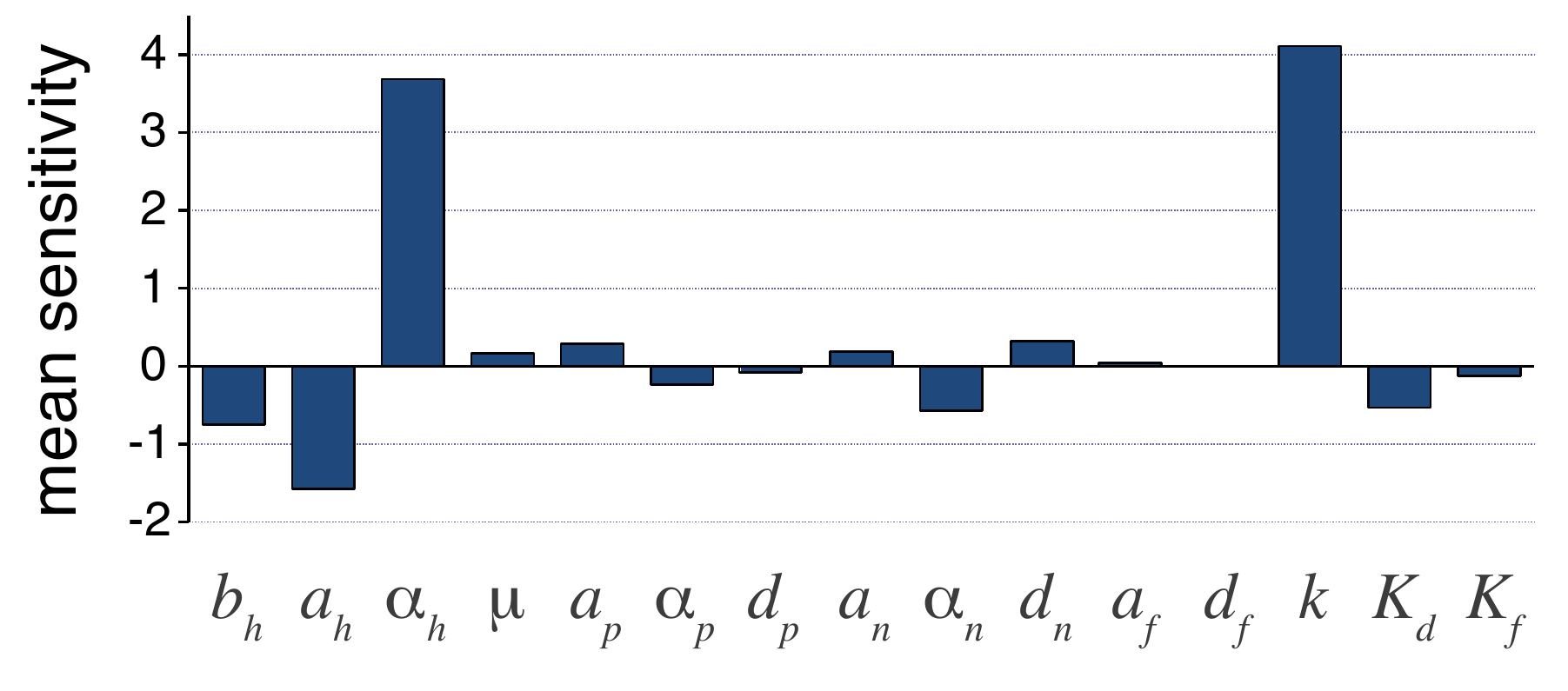}
\caption{Sensitivity of the mean distance between heterocysts after 96 hours with respect to $10\%$ changes in the indicated parameter values. This analysis is computed at the same point in parameter space employed to describe the wild-type phenotype, Table~S1.% \ref{tab.parameters}. 
\label{Fig.sensitivity} }
\end{figure*}

We have studied the model sensitivity and robustness by performing multiple simulations for all the parameters in our model and checking how the parameters affect the main pattern characteristics. 
We show the results regarding the mean distance between heterocysts, as one of the most relevant features of the heterocyst pattern (Fig. \ref{Fig.sensitivity}).
We find that the model is most sensitive with respect to {\em hetR} production and degradation rates, and its affinity to promoters. The model is very robust to small variations of any other parameter. This shows that the local positive feedback is the master regulator of the process, and we have used this fact to fit the model to experimental results using the three key parameters as the main handles to pattern properties. 
%

%\bibliographystyle{pnas}
%\bibliography{bib_anabaena}

\newpage


\newpage

\newpage

\begin{table*}
 \begin{minipage}{\textwidth}
 	\centering	 
	\caption{Parameters employed in wild-type simulations. Values in parentheses for $V^{-1}$ and $G$ are those employed for the weak and strong noise conditions (Fig.~7 of the main text and Fig.~\ref{fig.mean_noise}). %in Eqs.~ (\ref{Eq.HetR})-(\ref{function})
	} \label{tab.parameters}
	\begin{tabular}{@{} clc @{} clc @{}} % Column formatting, @{} suppresses leading/trailing space
		%\multicolumn{2}{c}{} \\
		Parameter 	&~ Description						&~ Value    	&~ 	Units	\\
		\hline
		$b_h$		&~HetR basal production rate				&~ 37.5 	&~	nM h$^{-1}$	\\
		$a_h$		&~HetR maximum regulated production rate 		&~ 1500 	&~	nM h$^{-1}$	\\
		$\alpha_h$	&~HetR degradation rate					&~ 4 		&~	h$^{-1}$	\\
		$\mu$		&~HetR dimer mediated degradation rate			&~ 0.001 	&~	nM$^{-1}$	\\
		$b_p$		&~PatS basal production rate				&~ 0 		&~	nM h$^{-1}$	\\
%		$b_p$		&~							&~ 5 		&~	nM/h	\\
		$a_p$		&~PatS  maximum regulated production rate		&~ 3000		&~	nM h$^{-1}$	\\
		$\alpha_p$	&~PatS degradation rate					&~ 2 		&~	h$^{-1}$	\\
		$d_p$		&~PatS  diffusion rate					&~ 10 		&~	h$^{-1}$	\\
		$b_n$		&~HetN basal production rate				&~ 0 		&~	nM h$^{-1}$	\\
%		$b_n$		&~ 							&~ 10 		&~	nM/h	\\
		$a_{n}$		&~HetN production rate in heterocysts			&~ 12000 		&~	nM h$^{-1}$	\\
		$\alpha_n$	&~HetN degradation rate					&~ 2.5		&~	h$^{-1}$	\\
		$d_n$		&~HetN diffusion rate					&~ 7.5		&~	h$^{-1}$	\\
		$b_f$		&~fixed nitrogen basal production rate			&~ 0 		&~	nM h$^{-1}$	\\
		$a_{f}$		&~fixed nitrogen production rate in heterocysts		&~ 1000	&~	nM h$^{-1}$	\\
		$\alpha_f$		&~fixed nitrogen degradation rate		&~ 2.5		&~	h$^{-1}$	\\
		$d_f$		&~fixed nitrogen diffusion rate				&~ 150	&~	h$^{-1}$	\\
		$k$			&~Affinity of HetR dimer to promoter		&~ 100 		&~	nM	\\
		$K_d$		&~Affinity of RGSGR to HetR				&~ 227	(ref. 41)&~	nM	\\
		$K_f$		&~Effective affinity of fixed nitrogen to HetR		&~ 227	&~	nM	\\
		$\rho$		&~Cellular growth rate					&~ 0.05 	&~	$\mu$m h$^{-1}$	\\
%		$V$			&~Effective volume setting noise strength in gene expression &~ 1.23		&~ nM$^{-1}$		\\
		$V^{-1}$	&~Noise strength in gene expression 			&~ 0.8 (0.05, 5)		&~ Dimensionless		\\
		$G$		&~Variance of noise in cellular growth			&~ 2.5 (0.6, 5.5)$\times10^{-4}$ 	&~ $\mu$m h$^{-1}$		\\
		\hline
	\end{tabular}
	%  \caption{Parameters of the model, Eqs.~(\ref{eq.full1}-\ref{eq.full9}).}
%\end{table*}
 \end{minipage}

 \begin{minipage}{\textwidth}
%\begin{table*}[H]
	\centering	 
	\caption{Stochastic terms employed wild-type in simulations. Values in parentheses are those employed for the weak and strong noise conditions (Fig.~7 of the main text and Fig.~\ref{fig.mean_noise}).} \label{tab.noise}
	\begin{tabular}{@{} clc @{} clc @{} clc @{}} % Column formatting, @{} suppresses leading/trailing space
		%\multicolumn{2}{c}{} \\
		Parameter 	&~ Noise description							&~ Mean    	&~ 	Standard deviation	&~ Units	\\
		\hline	
%		$\eta_{xi}(t)$	& Protein concentration dynamics					& 0	 	&	0.9 			& nM$^{1/2}$	\\
%		$\eta_{si}(t)$ 	& Growth process							& 0 		&	0.026			&($\mu$m h)$^{-1/2}$\\ %sqrt(Growth rate)*Noise_growth_rate
		$\eta_{di}$	& Maximum size for cell division					& 4 		&	0.42 (0.21, 0.84)	& $\mu$m	\\
		$\eta_{ci}$	& Maximum cumulative level for differentiation				& 960 		&	192 (64, 576)		& nM h				\\
		$\eta_{rs}$	& Minimum threshold level for differentiation reset			& 60		&	12 (4, 36)		& nM				\\
		\hline
	\end{tabular}
	%  \caption{Parameters of the model, Eqs.~(\ref{eq.full1}-\ref{eq.full9}).}
 \end{minipage}
\end{table*}

\end{document}